\documentclass[10pt,preprint]{aastex}

\newcommand\kms           {km~s$^{-1}$}

\newcommand{\tnm}         {\tablenotemark}
\newcommand{\tnt}         {\tablenotetext}
\newcommand{\hii}         {\ion{H}{2}}
\newcommand{\rarr}        {\ensuremath{\rightarrow}}

\begin{document}
\shorttitle{Full-Pol.\ OH in Massive SFRs: Results}

\title{Full-Polarization Observations of OH Masers in Massive
Star-Forming Regions: II.~Maser Properties and the Interpretation of
Polarization}
\author{Vincent L.~Fish\altaffilmark{1}}
\affil{Harvard--Smithsonian Center for Astrophysics}
\affil{60 Garden Street, Cambridge, MA  02138}
\affil{National Radio Astronomy Observatory}
\affil{P. O. Box O, 1003 Lopezville Road, Socorro, NM  87801}
\email{vfish@nrao.edu}
\and
\author{Mark J.~Reid}
\affil{Harvard--Smithsonian Center for Astrophysics}
\affil{60 Garden Street, Cambridge, MA  02138}
\email{reid@cfa.harvard.edu}

\altaffiltext{1}{Jansky Fellow}

\begin{abstract}

We analyze full-polarization VLBA data of ground-state, main-line OH
masers in 18 massive star-forming regions previously presented in a
companion paper.  From the aggregate properties of our sources, we
confirm results previously seen in the few individual sources studied
at milliarcsecond angular resolution.  The OH masers often arise in
the shocked neutral gas surrounding ultracompact \hii\ regions.
Magnetic fields as deduced from OH maser Zeeman splitting are highly
ordered, both on the scale of a source as well as the maser clustering
scale of $\sim 10^{15}$~cm.  Results from our large sample show that
this clustering scale appears to be universal to these masers.  OH
masers around ultracompact \hii\ regions live $\sim 10^4$ years and
then turn off abruptly, rather than weakening gradually with time.
These masers have a wide range of polarization properties.  At one
extreme (e.g., W75~N), $\pi$-components are detected and the
polarization position angles of maser spots show some organization.
At the other extreme (e.g., W51 e1/e2), almost no linear
polarization is detected and total polarization fractions can be
substantially less than unity.  A typical source has properties
intermediate to these two extremes.  In contrast to the well ordered
magnetic field inferred from Zeeman splitting, there is generally no
clear pattern in the distribution of polarization position angles.
This can be explained if Faraday rotation in a typical OH maser source
is large on a maser amplification length but small on a single
($e$-folding) gain length.  Increasing or decreasing Faraday rotation
by a factor of $\sim 5$ among different sources can explain the
observed variation in polarization properties.  Pure $\pi$-components
(in theory 100\% linearly polarized) have long been sought, but seldom
found.  We suggest that almost all $\pi$-components acquire a
signficant amount of circular polarization from low-gain stimulated
emission of a $\sigma$-component from OH appropriately shifted in
velocity and lying along the propagation path.

\end{abstract}


\section{Introduction}

Hydroxyl (OH) masers are common in massive star-forming regions
(SFRs).  Their small size and large Zeeman splitting coefficient allow
them to serve as probes of the velocity and magnetic fields on a very
small scale.  Because OH masers often cluster together in large
numbers on subarcsecond scales \citep[e.g.,][]{reidw3}, very long
baseline interferometric (VLBI) techniques are required to identify
individual maser features.  Due both to this close clustering of maser
spots and to the tendency for the two components of a ground-state
Zeeman pair to have highly unequal fluxes \citep{cook}, the resolution
afforded by VLBI is necessary in order to identify most Zeeman pairs
\citep[see, for instance,][]{fish03}.  Likewise, identifying
individual maser spots in a cluster is a prerequisite to understanding
the linear polarization of OH masers, since blending of the linear
polarization from adjacent masers with different polarization
properties can corrupt the interpretation of the polarization.

Over a quarter century has passed since the first OH maser source
was observed with VLBI resolution: W3(OH) by \citet{reidw3}.
Since then, only a few more interstellar ground-state OH maser sources
have been observed at VLBI resolution
\citep{haschick81,zheng97,slysh01b,slysh}.  In order to understand the
range of environments probed by OH masers in massive SFRs, we have
undertaken a survey of the OH masers in 18 massive SFRs with the
National Radio Astronomy Observatory's\footnote{The National Radio
Astronomy Observatory is a facility of the National Science Foundation
operated under cooperative agreement by Associated Universities, Inc.}
Very Long Baseline Array (VLBA).  The data have already been published
as \citet[hereafter Paper I]{paperi}.

In this paper we analyze the results both in terms of relevance to
individual sources and as a large collection of OH masers that can
shed additional light on the physical processes of OH masers and the
interpretation of their polarization.  A brief overview of linear
polarization theory is provided in \S \ref{lintheory}.  In \S
\ref{results}, we consider the set of OH masers as a whole to derive
statistical properties.  In \S \ref{discussion}, we present
plausibility arguments to attempt to identify the physical processes
responsible for the observed properties of OH masers.  Finally, we
summarize our results in \S \ref{conclusions}.  Comments on
individual sources are included in Appendix \ref{sourcenotes}.

\section{Linear Polarization Theory\label{lintheory}}

In the presence of a magnetic field, each of the main-line
$^2\Pi_{3/2}, J = 3/2$ OH transitions split into three lines: two
elliptically-polarized $\sigma$-components shifted in frequency by the
Zeeman effect, and one linearly-polarized $\pi$-component at the
natural frequency of oscillation for the velocity of the emitting
material \citep[see, e.g.,][]{davies}.  Due in
part to unequal amplification, this clear pattern of three lines has
only been seen once \citep{hutawarakorn}.  To this point, there have
been few, if any, unambiguous detections of $\pi$-components.  Often,
a single $\sigma$-component is seen by itself, without an accompanying
$\sigma$-component polarized in the opposite circular handedness
strong enough to be detected.  Significant linear polarizations, which
could be produced by $\pi$-components, have been seen \citep{slysh}.
However, $\sigma$-components associated with magnetic fields with a
nonzero perpendicular (line-of-sight) component are in general
elliptically polarized, and elliptical polarization is the sum of
circular and linear polarizations.  Thus, significant linear
polarization fractions can be produced by both $\sigma$- and
$\pi$-components, the latter theoretically being 100\% linearly
polarized in all instances.

The existence of maser spots with large (but not unity) fractional
linear polarization suggests that $\pi$-components may in practice be
partially circularly polarized as well.  It is difficult to select an
observational threshold of the linear polarization fraction that
divides $\sigma$-components from $\pi$-components.  Theoretical
modelling by \citet{gkk3} and \citet{grayfield} suggests that
amplification of $\pi$-components is stronger than that of
$\sigma$-components for $\theta \gtrsim 55\degr$, where $\theta$ is the
angle of the magnetic field to the line of sight.  This corresponds to
a linear polarization fraction $\geq 0.50$.  But since linear and
circular polarization fractions add in quadrature, a completely
polarized spot is not more linear than circular until the linear
polarization fraction exceeds 0.71.  Even then it is unclear how
$\pi$-components, which are theoretically totally linearly polarized,
can acquire a circular component of polarization.  This issue
is discussed in further detail in \S \ref{overlap}.

It is important to distinguish between $\sigma$- and $\pi$-components,
because the interpretation of the orientation of the magnetic field
based on the linear polarization position angle is different for the
two cases.  The electric vector for a $\sigma$-component is
perpendicular to the two-dimensional magnetic field direction (i.e.,
on the plane of the sky), while the electric vector for a
$\pi$-component is parallel to the two-dimensional magnetic field
direction.  Thus, without information as to which maser spots are
$\sigma$-components and which are $\pi$-components, there is a
$90\degr$ ambiguity in the direction of the magnetic field on the
plane of the sky.

\section{Results\label{results}}

\subsection{Are Zeeman Triplets Ever Seen?\label{triplet}}

\begin{figure}
\begin{center}
\includegraphics[width=6.0in]{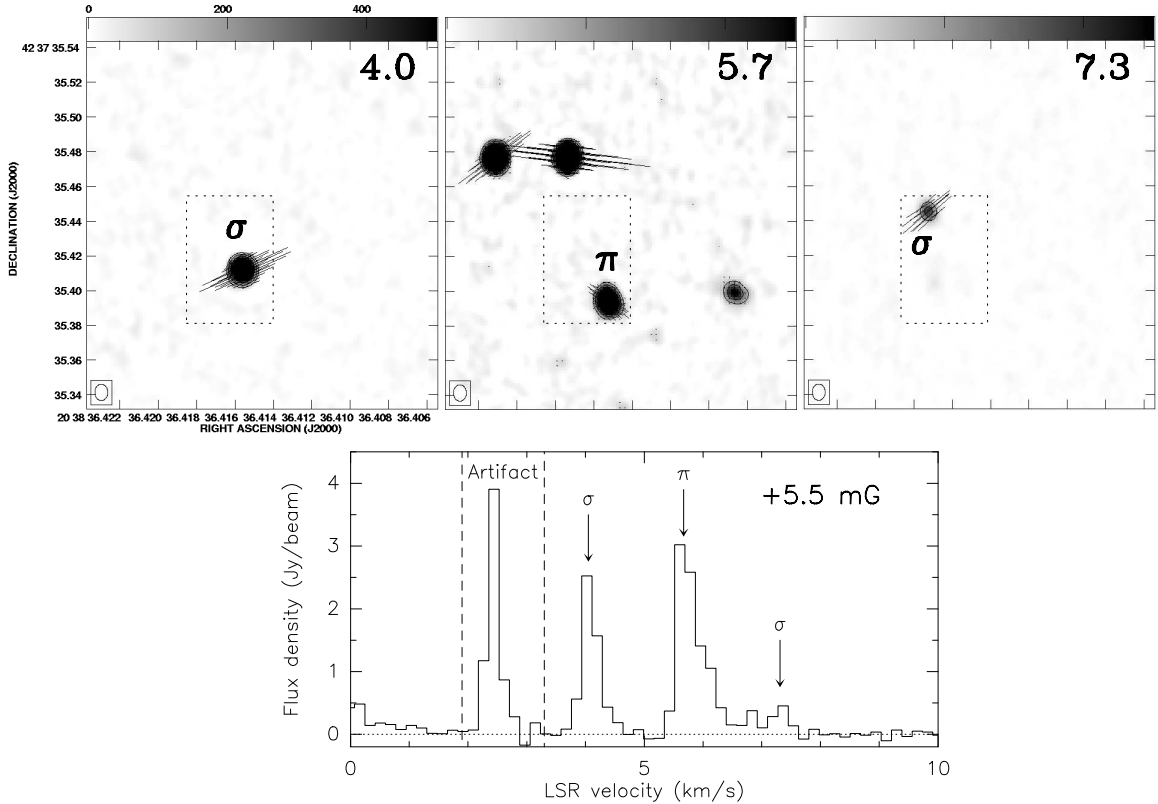}
\end{center}
\singlespace
\caption[Zeeman triplet in W75 N]{Zeeman triplet in W75 N.  Top: Contour
maps of the three Zeeman components with polarization vectors
included.  Note that the polarization vectors of the
$\sigma$-components are roughly perpendicular to that of the
$\pi$-component.  LSR velocities in \kms\ are shown in the upper right
of each panel.  Bottom: Spectrum of the dotted box region in the upper
plots.  The feature labelled ``artifact'' is due to the sidelobe of a
very strong maser spot outside the region shown.  The velocities of
the three marked components are consistent with a $+5.5$~mG magnetic
field.\label{w75n-triplet}}
\doublespace
\end{figure}

\citet{hutawarakorn} identified a complete Zeeman triplet in the
northernmost group of OH masers in \object[W75N]{W75~N}, the first and
only OH Zeeman triplet identified.  Figure \ref{w75n-triplet} shows a
spectrum of a subregion of the northernmost group of OH masers.  Three
maser lines can be seen at 4.1, 5.7, and 7.3~\kms.  The splitting of
the 4.1 and 5.7~\kms\ maser lines is consistent with a magnetic field
of $+5.5$~mG and a velocity of 5.7~\kms.  We confirm the existence of
this triplet, although the three components are not perfectly aligned
spatially.  The angular separation between the $\pi$- and RCP
$\sigma$-component is 55 mas, which corresponds to a two-dimensional
linear separation of $1.6 \times 10^{15}$~cm (110 AU).  This is
roughly the cluster scale of OH masers (see \S \ref{clusterscale}),
though larger than the typical scale of separation between the two
$\sigma$-components in a Zeeman pair.  It is possible that the $\pi$-
and both $\sigma$-components all are located in the same cloudlet but
from amplification paths that are not coincident.

The linear polarization fractions of the $\sigma$-components are 16\%
and 20\%, while the corresponding fraction of the $\pi$-component is
86\%.  The polarization position angles of the $\sigma$-components at
4.1 and 7.3~\kms\ are $-45\degr$ and $-67\degr$, respectively, while
the position angle of the $\pi$-component at 5.7~\kms\ is $54\degr$.
This roughly fits the theory that $\pi$-components are linearly
polarized with a position angle perpendicular to that of the
$\sigma$-components.

The angle of the magnetic field to the line of sight, $\theta$, can be
derived from the formula
\begin{equation}
f = \frac{\sin^2 \theta}{1 + \cos^2\theta},
\end{equation}
where $f$ is the linear polarization fraction of a $\sigma$-component
\citep{gkk2}.  This angle is $32\degr$ and $35\degr$ for the two
$\sigma$-components of the Zeeman triplet.  Equal amplification rates
of the $\sigma$- and $\pi$-components occur when $\sin^2 \theta =
2/3$, or for $\theta \approx 55\degr$, with $\pi$-components favored
when the magnetic field is more highly inclined to the line of sight
\citep{gkk3}.  Modelling of OH hyperfine populations for conditions
typically found at maser sites shows that the amplification of
$\pi$-components falls off rapidly for $\theta < 55\degr$
\citep{grayfield}.  For smaller $\theta$, beaming and competitive gain
favor $\sigma$-components, suppressing amplification of
$\pi$-components.  Yet the inclination of the magnetic field to the
line of sight is $\sim 35\degr$, not $55\degr$, in the only confirmed
Zeeman triplet.

\begin{figure}
\begin{center}
\includegraphics[width=6.0in]{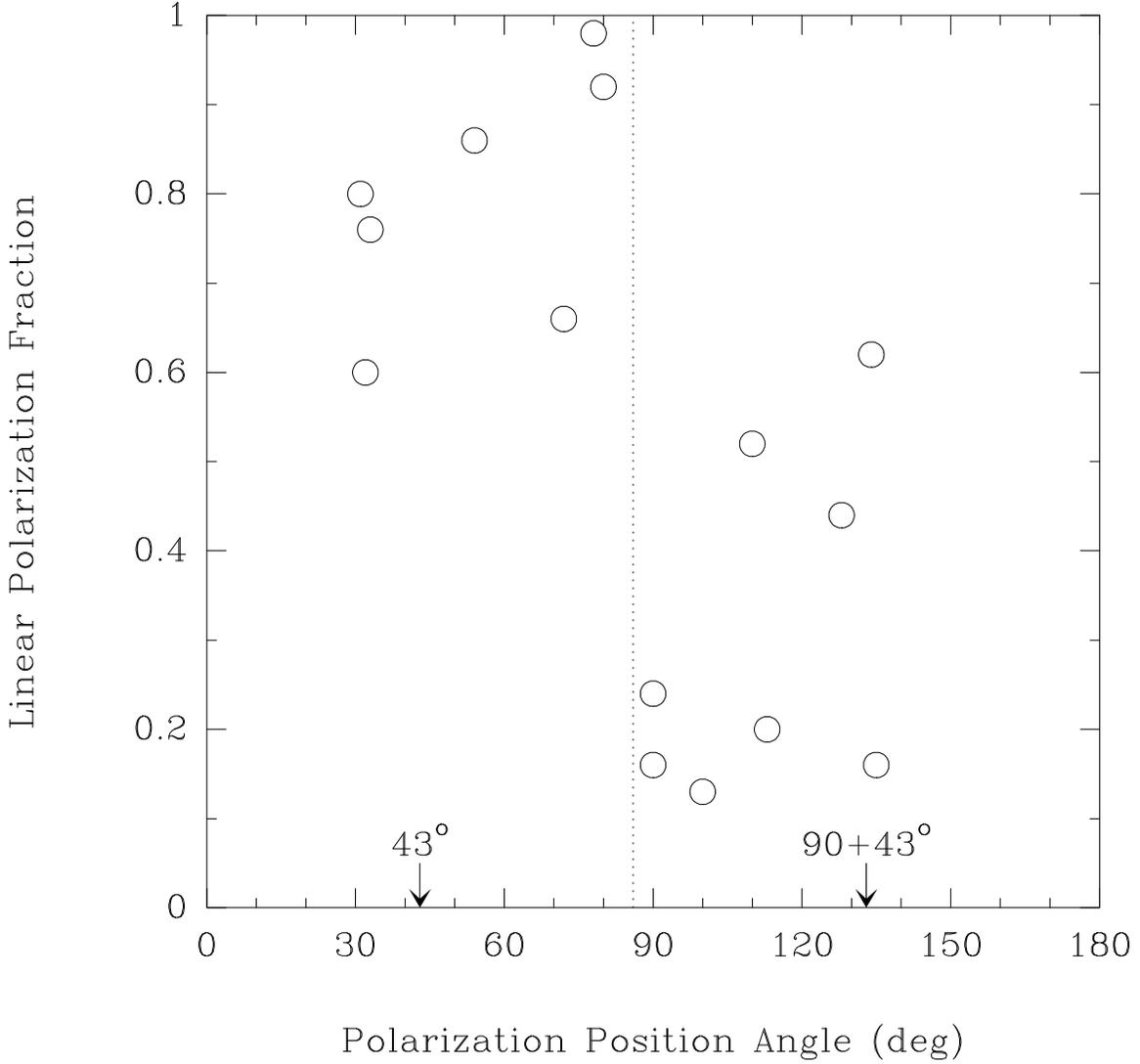}
\end{center}
\singlespace
\caption[Linear polarization fractions in northern cluster of W75
N]{Linear polarization fractions as a function of polarization
position angle (PPA) in the northern cluster of W75 N.  The elongation
angle of VLA~1 (43$\degr$) is marked along with the angle
perpendicular to it.  The PPAs of highly linearly-polarized spots
(left of dotted line) tend to be aligned parallel to the axis of the
continuum source, while the PPAs of highly circularly-polarized spots
(right of dotted line) are roughly oriented perpendicular.  If these
groups are interpreted as $\pi$- and $\sigma$-components respectively,
the implied projected magnetic field direction is at a position angle
near $43\degr$.\label{w75n-ppa}}
\doublespace
\end{figure}

Figure \ref{w75n-ppa} shows a plot of the fractional linear
polarization of maser spots in the northernmost cluster of OH spots in
W75~N versus the position angle of polarization.  The masers seem to
be grouped into two populations.  The first group consists of masers
whose position angle is less than $90\degr$.  These masers have a high
linear polarization fraction.  The second group, with position angles
$> 90\degr$, are mostly circularly polarized.  If we interpret these
groups as $\pi$- and $\sigma$-components respectively, the magnetic
field implied from linear polarization lies at position angle
$48\degr$ in the region, with a scatter of $24\degr$.  This would
indicate that the direction of the magnetic field on the plane of the
sky is roughly aligned along the NE-SW distribution of maser spots and
the elongation of the continuum source VLA~1.

\subsection{Clustering Scale\label{clusterscale}}

\begin{figure}
\begin{center}
\includegraphics[width=6.0in]{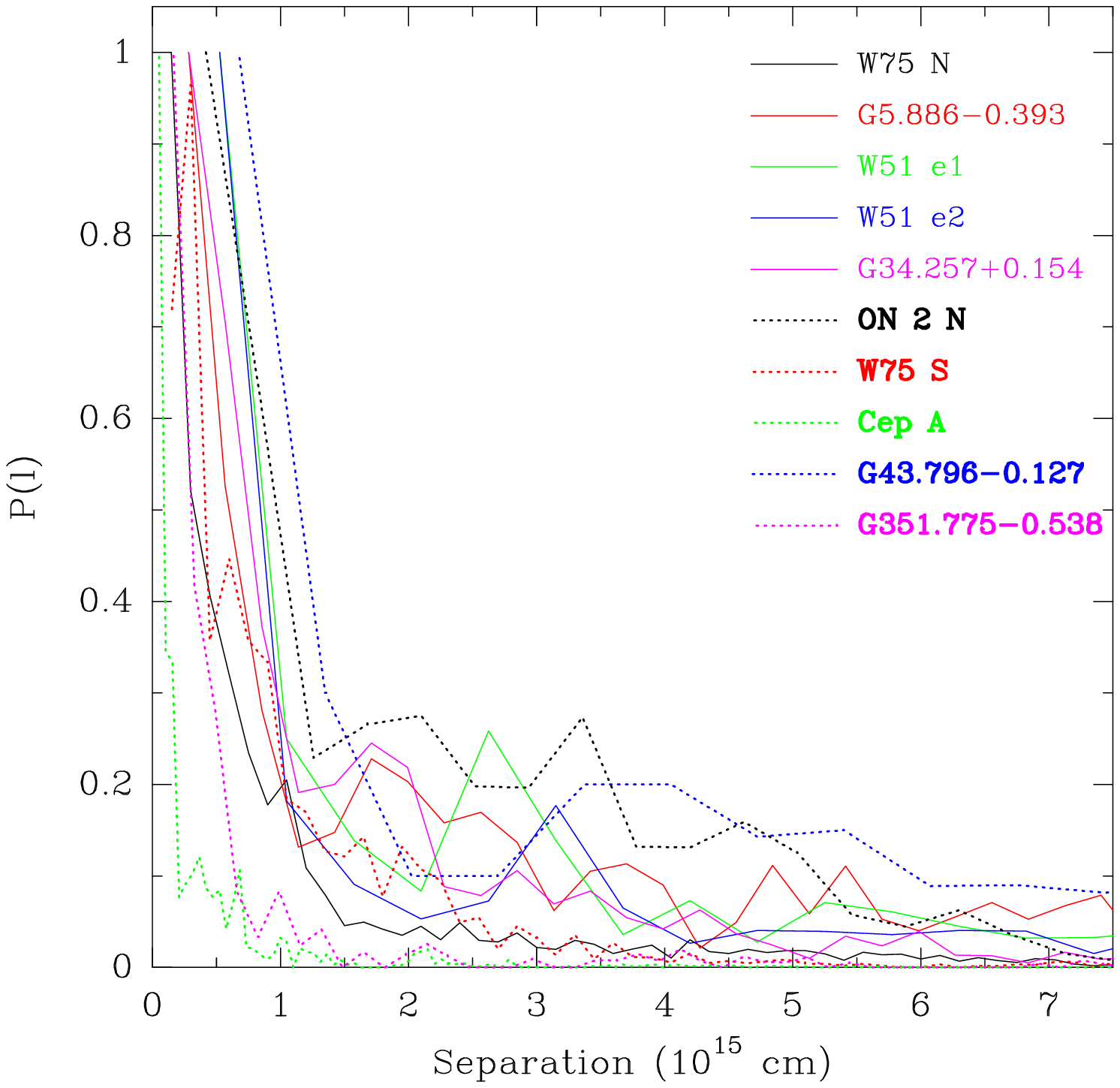}
\end{center}
\singlespace
\caption[Two-point spatial correlation function of masers in massive
SFRs.]{Two-point spatial correlation function of the masers in ten
massive SFRs.  $P(l)$ is the probability per unit solid angle that a
maser spot can be found at angular distance $l$ from any given maser
spot.  The angular distances have been multiplied by the distance to
each source to give the linear separation, plotted as the abscissa.
The probabilities $P(l)$ have been normalized to the greatest value.
Note that $P(l)$ drops to $e^{-1}$ at a separation of about
$10^{15}$~cm for all ten sources.\label{fig-all-seps}}
\doublespace
\end{figure}

In W3(OH), \citet{reidw3} found that the clustering scale of OH masers
was approximately $10^{15}$~cm.  They calculated a two-point spatial
correlation function and found that the probability per unit solid
angle of finding another maser spot within angular distance $l$ of a
spot decreased sharply for $l \gtrsim 10^{15}$~cm.  We performed a
similar analysis on each of our sources, and the results for the ten
having a large number of maser spots are shown in Figure
\ref{fig-all-seps}.  For all ten sources, the probability $P(l)$ drops
sharply between a separation of 0 and $10^{15}$~cm (67 AU).  The other
sources in this study show similar behavior, although the plots are
``noisier'' owing to the smaller number of maser spots in the sources.
This evidence argues in favor of a common clustering scale for OH
masers in all massive SFRs.  The maser spots within these clusters
represent paths in the same condensation where the physical conditions
are favorable to exponential amplification.  If so, the overall extent
of these condensations is perhaps a few times as large ($\sim
150$~AU), since paths through the periphery of the condensation are
much less likely to produce comparable amplification lengths.  In \S
\ref{shocks} we suggest that instabilities in the shocked neutral gas
may lead to the formation of such condensations.

\subsection{Zeeman Pairs and Component Intensities\label{components}}

\begin{figure}
\begin{center}
\includegraphics[width=5.0in]{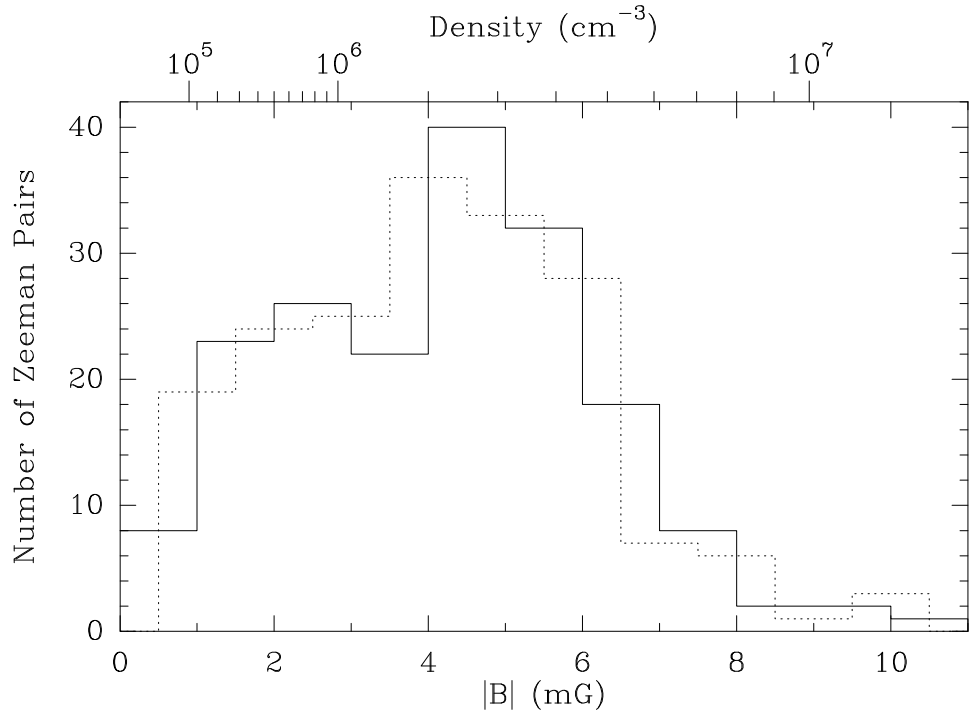}
\end{center}
\singlespace
\caption[Histogram of magnetic field strengths.]{Histogram of
magnetic field strengths implied by Zeeman splitting.  The solid and
dotted lines plot the same data but with the bins shifted by 0.5~mG.
The dropoff near zero is due to observational limits.  The implied
total density is shown at the top, assuming $n_{\mathrm H_2} = 2 \times
10^{6}$~cm$^{-3}$ at 4~mG as suggested for W3(OH) \citep{rmb} and an
$n \propto B^2$ scaling law.  The 19.8 and 21.0~mG fields in W51,
not included in this plot, would correspond to a density of $5 \times
10^7$~cm$^{-3}$ under these assumptions.\label{zeeman-numbers}}
\doublespace
\end{figure}

In total, we find 184 Zeeman pairs in the entire sample set.  These
Zeeman pairs are listed in Table 21 of Paper I.  A histogram of
the distribution of implied magnetic field strengths is given in
Figure \ref{zeeman-numbers}.  The distribution rises with increasing
magnetic field strength to about 4~mG, then falls.

Few Zeeman pairs are found that imply a splitting greater than about
8~mG, and the largest magnetic field strength found is 21~mG in
\object{W51} e2.  There is theoretical support for the existence of an
upper limit to the strength of the magnetic field in a Zeeman pair for
OH masers.  The collapse of material increases both the density and
the magnetic field; \citet{mouschovias} suggests that the relation
could be as steep as $n \propto B^2$, and Zeeman measurements of the
magnetic field in molecular clouds are consistent with this relation
\citep[e.g.,][]{crutcher}.  At some density, the rate of collisional
deexcitation will be higher than the pump rate (presumably from
radiative excitation), and the population inversion between the lower
and upper states will be destroyed.  The rate of collisions of H$_2$
with OH is $n_{{\mathrm H}_2} \langle\sigma v\rangle$, where
$\langle\sigma v\rangle = 10^{-9} - 10^{-10}$~cm$^{3}~$s$^{-1}$, and
the pump rate likely is $\sim 0.03$~s$^{-1}$, as would be expected for
a far-infrared rotational transition \citep{goldreichhouches}.  Thus
the population inversion will be destroyed by thermal collisions when
the density is a few times $10^8$~cm$^{-3}$.  We note that the range
of detected magnetic fields in our Zeeman pairs is a factor of 35 (0.6
to 21 mG), or roughly $\sqrt{10^3}$.  If the $B^2 \propto n$ scaling
law applies throughout this entire range, this implies that the range
of densities sampled by those masers in Zeeman pairs is a factor of
$10^3$, or about $n_{{\mathrm H}_2} = 10^5 - 10^8$, assuming that the
21~mG magnetic field in W51 e2 is near the upper end of magnetic field
strengths allowable before collisional depopulation of the upper state
is significant.

\begin{figure}
\begin{center}
\includegraphics[width=5.0in]{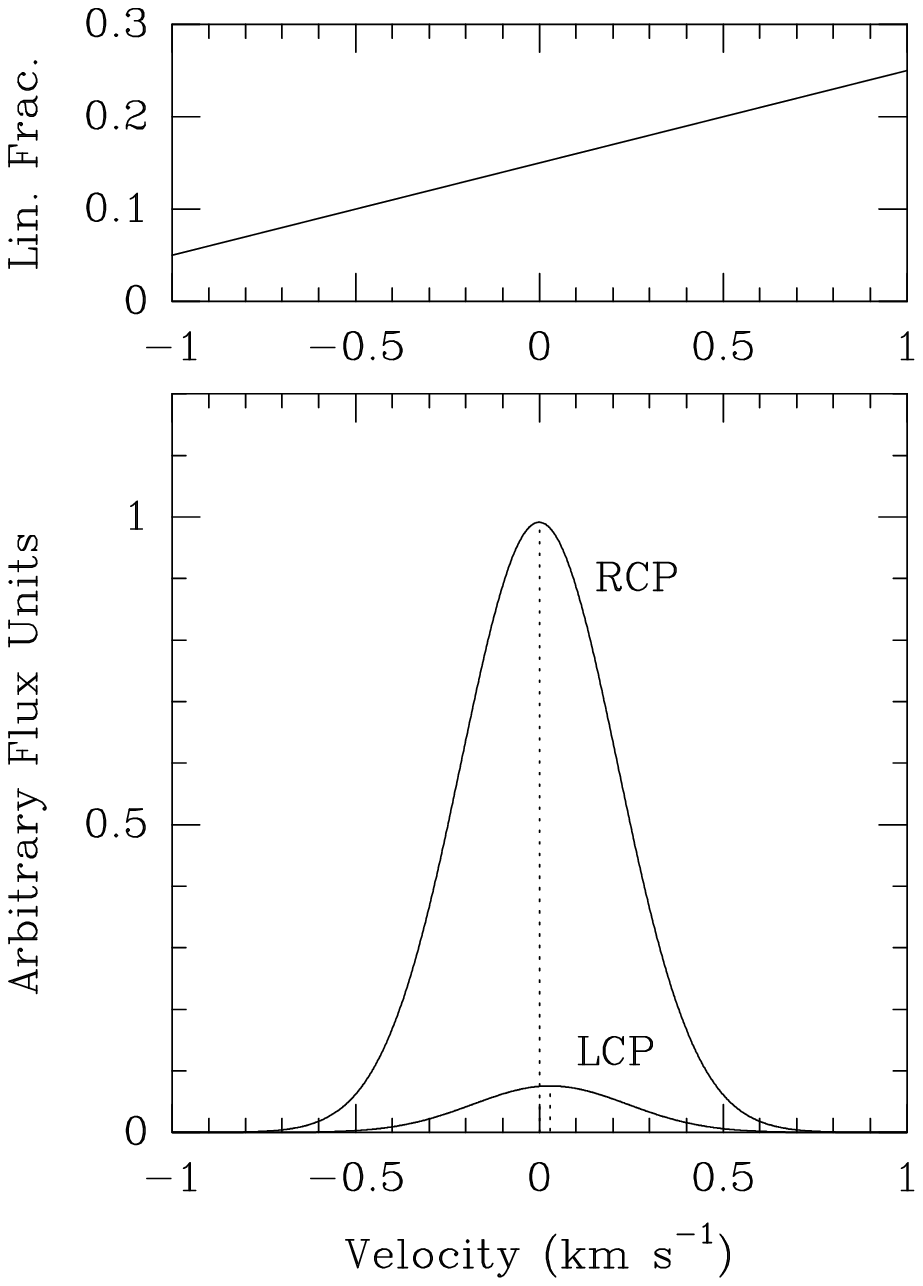}
\end{center}
\singlespace
\caption[Velocity shift due to a frequency-dependent linear
polarization fraction.]{Velocity shift due for a right
elliptically-polarized line with FWHM 0.5 \kms\ whose linear
polarization fraction varies across the linewidth as shown in the top
panel.  The linear fraction could vary in this manner (for instance)
due to a bend in the magnetic field along the amplification path.  The
bottom panel shows the line as would be detected by RCP and LCP feeds.
The LCP line is shifted 0.03 \kms\ with respect to the RCP
line.\label{freq-dep}}
\doublespace
\end{figure}

The lack of Zeeman pairs below 0.5~mG may be an observational effect.
For such small magnetic fields, the splitting is less than a typical
line-width, and we make no attempt to identify Zeeman pairs less than
0.5~mG due to the difficulty of distinguishing small Zeeman shifts
from other effects.  For example, consider a right-elliptically
polarized $\sigma$-component in a region where the magnetic field
orientation varies along the amplification path, as in Figure
\ref{freq-dep}.  The linear component of the polarization will be seen
as weak emission in LCP.  If there is also a velocity gradient along
the amplification path, the linear polarization component may be
shifted in velocity with respect to the circular polarization
component.  This would manifest itself in our observations as a
velocity difference between the lines seen in the two
circularly-polarized feeds: the RCP feed would detect nearly all the
emission, while the LCP feed would detect only the weaker,
velocity-shifted linear component.  For the parameters shown in Figure
\ref{freq-dep}, the shift between LCP and RCP velocity corresponds to
an apparent magnetic field strength of 0.1~mG at 1667 MHz in the
Zeeman interpretation, although broader lines and more extreme
variations of the linear polarization fraction across a linewidth can
produce larger apparent shift between the components.

\begin{figure}
\begin{center}
\includegraphics[width=5.0in]{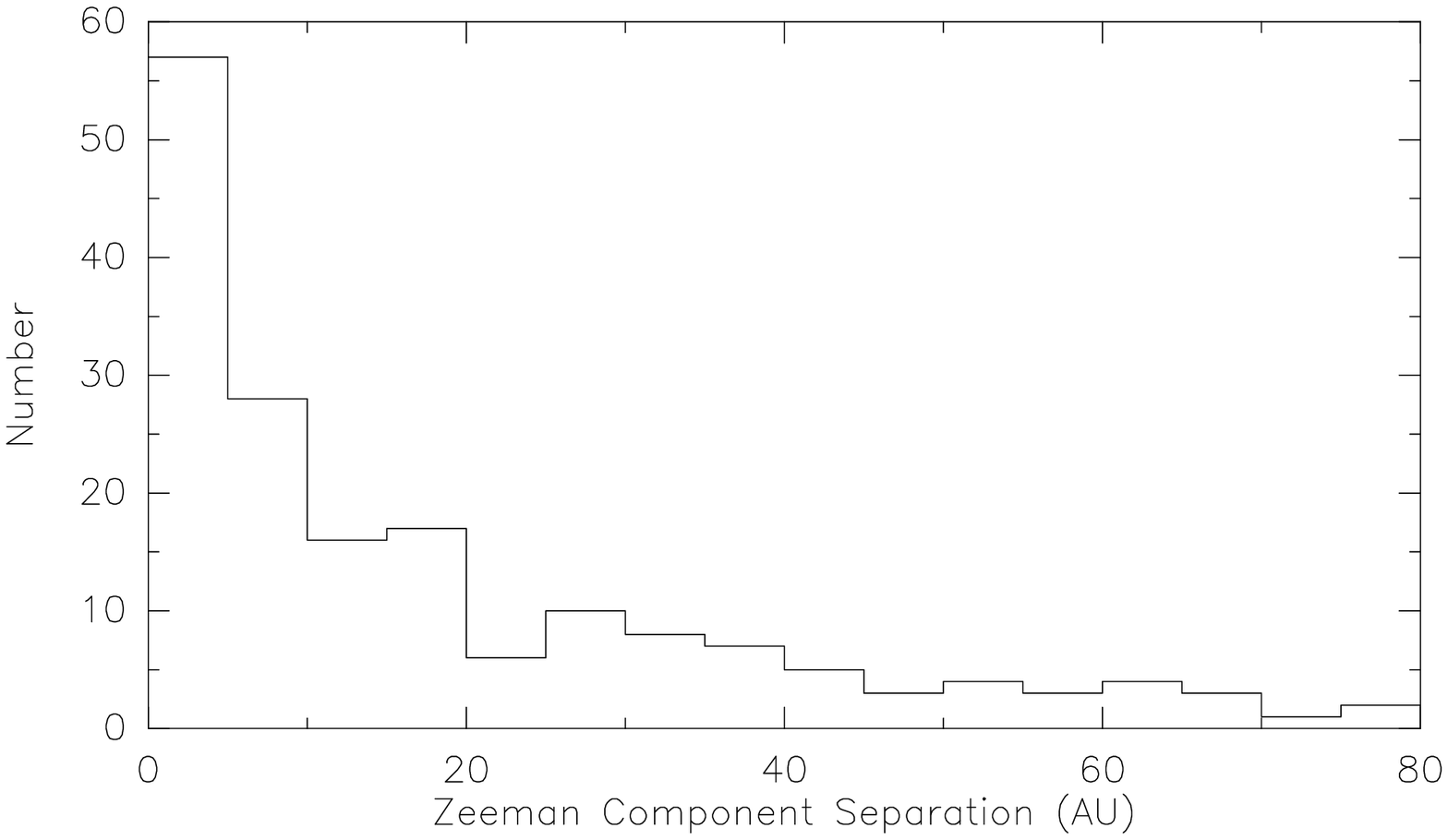}
\end{center}
\singlespace
\caption[Histogram of identified Zeeman component
separations.]{Histogram of identified Zeeman component separations.
Note the rapid falloff of the distribution of $\sigma$-component
separations in a Zeeman pair.  This suggests that the components of a
Zeeman pair are generally separated by a distance less than the
$10^{15}$~cm ($\sim 67$~AU) clustering scale (see \S
\ref{clusterscale}).  Ten Zeeman pairs with component separations
exceeding 80 AU are not shown.  The plot is not normalized by area;
the falloff would be much faster if it were. \label{zeemanseps}}
\doublespace
\end{figure}

Figure \ref{zeemanseps} shows a histogram of the separation between
the $\sigma$-components in each Zeeman pair for which the
$\sigma$-component separation is less than $\approx 10^{15}$~cm.
Zeeman pairs with larger separations are identified when unambiguous,
but they have been excluded from the present consideration in order
not to introduce bias.  The rapid falloff of the number of pairs
identified with increasing component separation suggests that the
dearth of identifiable Zeeman pairs at larger separations is real.
Note that Figure \ref{zeemanseps} has not been normalized by area;
were it to be so normalized, it would fall even faster.  If the
distribution of Zeeman component separations were uniformly random, a
plot of the unnormalized number of identifiable Zeeman components
versus component separation would be an \emph{increasing} function of
separation (at least up to a distance beyond which pairing is no
longer unambiguous).  This is certainly not observed, providing
further evidence that the $10^{15}$~cm clustering scale is a
physically significant scale over which physical parameters are
sufficiently coherent to provide an environment conducive to maser
activity.

Thus it appears that the spatial separation of the $\sigma$-components
in a Zeeman pair is generally a factor of several smaller than the
size of the cluster containing the pair.  Nevertheless, there are some
pairs of maser spots polarized predominantly in opposite circular
senses whose separation is comparable to or exceeds $10^{15}$~cm.  It
is possible that these are not true Zeeman pairs of
$\sigma$-components from the same masing subregion but rather two
oppositely-polarized $\sigma$-components from different Zeeman pairs,
of which only one component is seen in each.  Such a situation could
arise if amplification at each maser site favors only one sense of
circular polarization, as described below.  The magnetic fields implied
by these Zeeman ``cousins'' would be less accurate due to two
effects.  First, the systemic velocity at each maser site may be
different due to turbulence.  \citet{reidw3} calculate that a typical
intracluster turbulent velocity is 0.6~\kms.  This corresponds to an
effective Zeeman splitting of 1.0 mG in the 1665 MHz transition and
1.7 mG in the 1667 MHz transition.  Second, the magnetic field
strength may be different at the maser sites.  This difference is
generally less than 1 mG (see \S \ref{magstruc}).  The magnetic field
strengths implied by Zeeman cousins should therefore be accurate to
better than 2 mG.

\begin{figure}
\begin{center}
\includegraphics[width=5.0in]{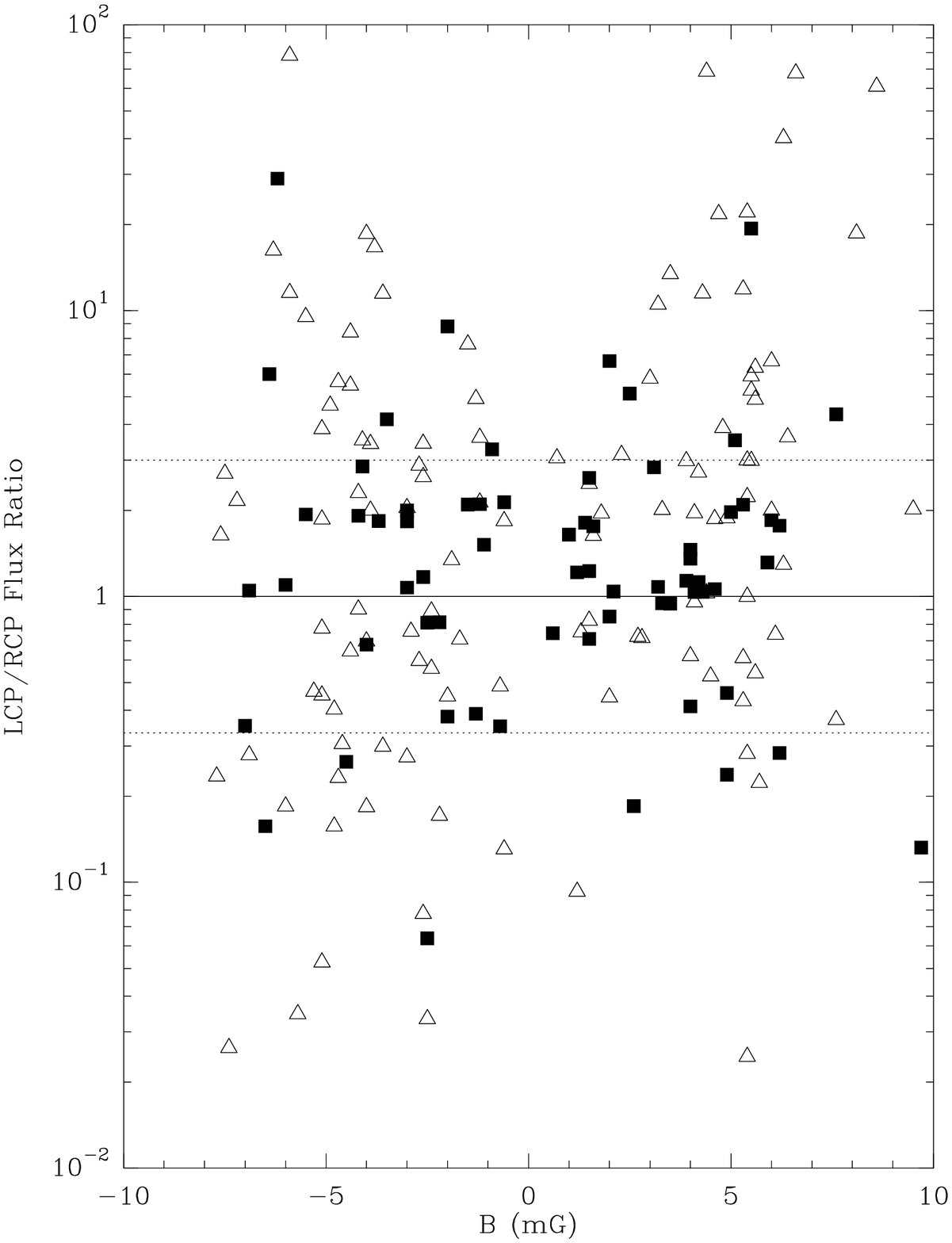}
\end{center}
\singlespace
\caption[Flux ratios of $\sigma$-components in Zeeman pairs.]{Flux
ratios of $\sigma$-components in Zeeman pairs.  Data for 1665 MHz
pairs are represented by open triangles, while data for 1667 MHz pairs
are represented by filled squares.  Dotted lines are drawn at ratios
of 3 and $1/3$.  The 19.8 and 21.0~mG fields in W51 are not
included.\label{fluxratios}}
\doublespace
\end{figure}

Figure \ref{fluxratios} shows the ratio of the LCP and RCP fluxes for
the $\sigma$-components in each Zeeman pair.  We observe that this
ratio occasionally reaches values near 100.  This is probably not a
hard upper limit but a result of observational constraints.  Maser
spots weaker than 50 to 100 mJy are too weak to be detected in our
survey; maser spots stronger than 10 Jy are rare.

A simple test to determine whether our data have unknown systematics
is to compare the number of Zeeman pairs with greater flux in the LCP
and RCP components.  There are 119 Zeeman pairs in the 1665 MHz
transition and 65 in the 1667 MHz transition.  The LCP component is
stronger than the RCP in 69 of the 1665 MHz pairs and 44 of the 1667
MHz pairs.  If Zeeman pairs with stronger LCP and RCP components are
equally common, we would expect the number of Zeeman pairs in each
polarization to be $59.5 \pm 5.5$ at 1665 MHz and $32.5 \pm 4.0$ at
1667 MHz\footnote{For a binomial distribution, $\sigma^2 = Npq$, where
$N$ is the sample size and $p$ and $q$ are the probabilities of each
outcome ($S_\mathrm{LCP} > S_\mathrm{RCP}$ and $S_\mathrm{LCP} <
S_\mathrm{RCP}$).}.  The deviations from these values are not
statistically significant.  The ratio of intensities of the
$\sigma$-components in a Zeeman pair does not appear to be
substantially different for the 1665 and 1667 MHz transtions.

\begin{figure}
\begin{center}
\includegraphics[width=5.0in]{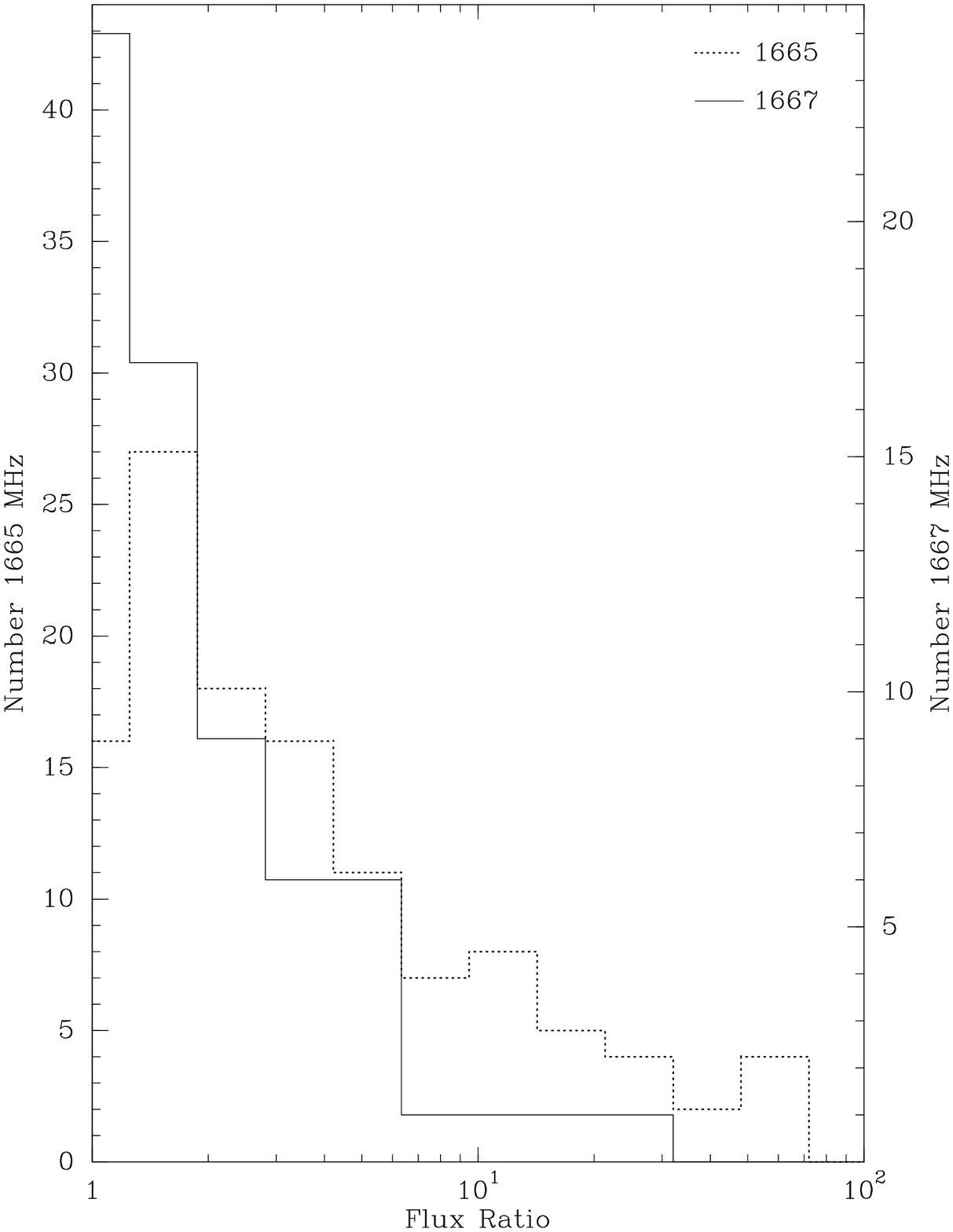}
\end{center}
\singlespace
\caption[Histogram of flux ratios in Zeeman pairs.]{Histogram of flux
ratios in Zeeman pairs.  The flux ratio, determined by taking the
ratio of the stronger flux to the weaker flux in a Zeeman pair, is
binned by factors of 1.5.  The dotted line shows data for the 1665~MHz
transition, and the solid line shows data for the 1667~MHz transition.
The data for the two transitions are scaled by the total number of
Zeeman pairs identified in the transition.  The histograms are similar
for flux ratios less than 10, but 25 of the 28 pairs with flux ratios
greater than 10 are in the 1665~MHz transition.\label{flux-ratios-freq}}
\doublespace
\end{figure}

\begin{figure}
\begin{center}
\includegraphics[width=5.0in]{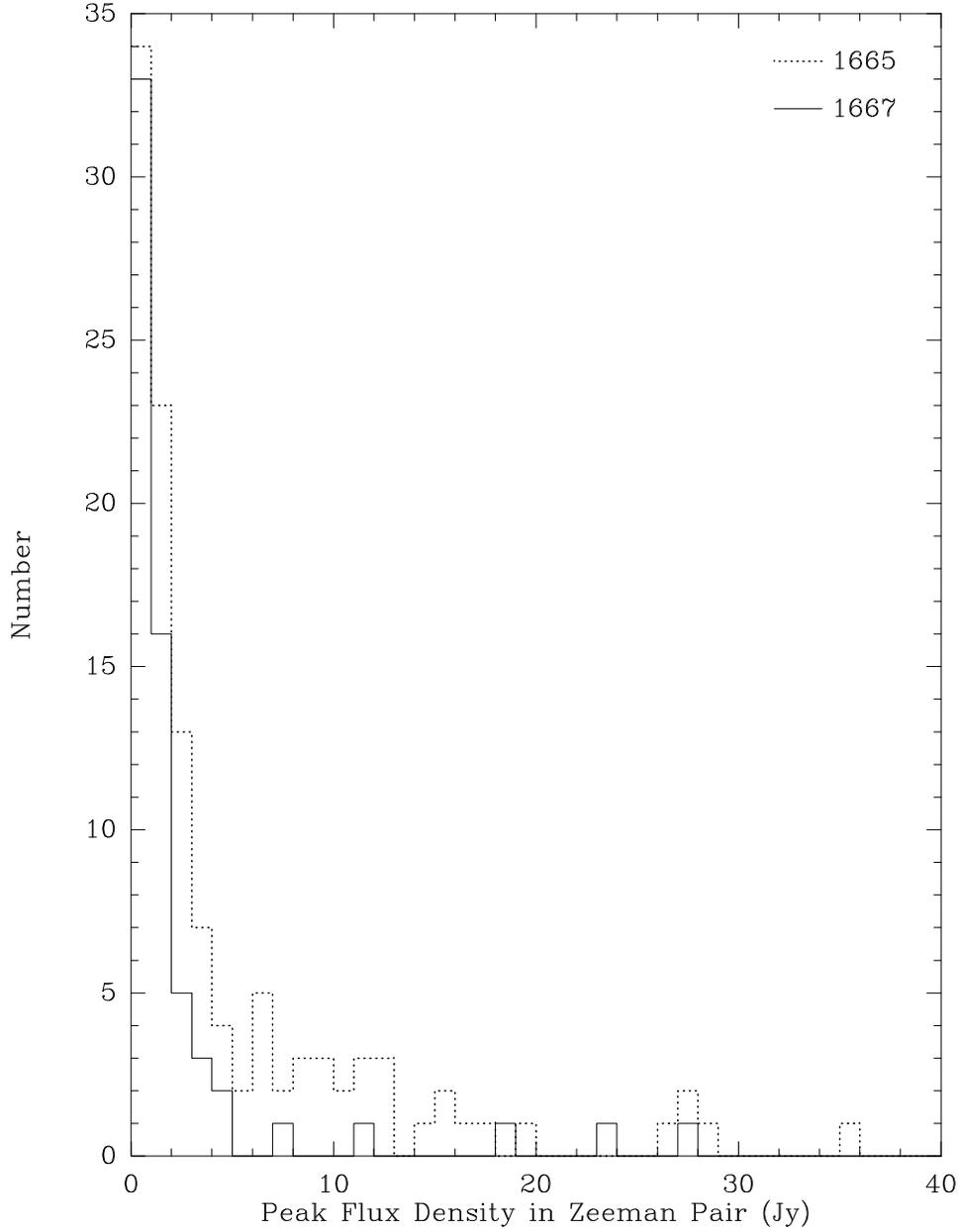}
\end{center}
\singlespace
\caption[Histogram of peak flux densities in Zeeman pairs.]{Histogram
of peak flux densities in Zeeman pairs (peak flux density in either
LCP or RCP).  Flux densities are binned by increments of 1 Jy.  For
clarity, points stronger than 40~Jy are suppressed.  These consist of
three points at 1665~MHz (72, 163, 239~Jy) and two at 1667~MHz (41,
85~Jy).  The stronger component of a 1665~MHz pair tends to be
stronger than that of a 1667~MHz pair.\label{flux-binning}}
\doublespace
\end{figure}

Figure \ref{flux-ratios-freq} shows a histogram of the flux ratios of
Zeeman components.  There is not an appreciable difference between the
1665 and 1667~MHz transitions for flux ratios less than 10.  The
larger number of high flux ratios ($> 10$) for the 1665~MHz transition
appears real, although it may partially be due to a selection effect.
As Figure \ref{flux-binning} shows, the brighter component in a
typical 1665~MHz Zeeman pair is brighter than that of a 1667~MHz pair.
The detection limit in our survey varies somewhat by source but is
approximately 0.1~Jy.  Thus, the identification of a Zeeman pair with
flux ratio $x$ requires that the stronger component have a flux
density higher than $0.1\,x$~Jy.  The relative scarcity of 1667~MHz
Zeeman pairs with a flux ratio greater than 10 can be explaned by the
paucity of Zeeman components with a flux density greater than 1~Jy.
Only in one-third of cases does a 1667~MHz Zeeman pair include a
component stronger than 1~Jy, while over half of 1665~MHz pairs
include a component above this threshold.  At higher flux density
thresholds, the difference becomes more extreme.

The relative absence of 1667~MHz Zeeman pairs with large flux ratios,
noted previously in W3(OH) \citep{wright04b}, is consistent with the
picture that maser transitions with smaller Zeeman splitting
coefficients tend to have Zeeman pairs in which the
$\sigma$-components are more equal in intensity
\citep[e.g.,][]{moran78, caswellv}.  \citet{cook} theorized that
correlated velocity and magnetic field gradients could be the cause of
unequal spot intensities in Zeeman pairs.  \citet{deguchi} argued that
even absent a magnetic field gradient, a velocity gradient alone is
sufficient to produce unequal intensities.  Velocity gradients (either
alone or in combination with magnetic field gradients) are less likely
to disrupt amplification of only one of the $\sigma$-components of a
Zeeman pair for the 1667~MHz transition than for the 1665~MHz
transition.  Measured line widths for maser spots at 1667 and 1665~MHz
are similar, but the Zeeman splitting coefficient is smaller for the
1667~MHz transition than for the 1665~MHz transition.  The magnetic
field strength required to produce an effective velocity shift greater
than the linewidth of a component is therefore greater for 1667~MHz
masers than for 1665~MHz masers.

\subsection{Maser Spot Statistics}

We find a total of 342 spots stronger in RCP and 351 in LCP at
1665~MHz, as well as 178 in RCP and 185 in LCP at 1667~MHz.  As
expected, there does not appear to be a preference for masers to
appear preferentially in one circular polarization than the other,
though occasionally an individual source may have a preponderance of
maser spots in one polarization, as is the case in W75 S, in which we
find 35 RCP and 19 LCP spots in the 1665~MHz transition.  For a source
with 54 spots, we would expect $27 \pm 4$ spots in each polarization,
so the deviation seen in W75 S is not significant.

There are nearly twice as many spots detected at 1665~MHz than at
1667~MHz, although one source
(\object[G40.622-0.137]{G40.622$-$0.137}) has one more spot in
1667~MHz than at 1665~MHz.  This fits with theoretical modelling,
which shows that while 1665~MHz and 1667~MHz masing often occur under
the same physical conditions, the area of physical parameter space
conducive to amplification is larger for 1665~MHz masers than for
1667~MHz masers \citep{cragg}.

\subsection{Magnetic Field Structure\label{magstruc}}

Magnetic fields as determined from Zeeman splitting of OH masers are
predominantly ordered in massive SFRs \citep[e.g.,][]{baart85,garcia}.
In all sources with the possible exception of
\object[G43.796-0.127]{G43.796$-$0.127}, the line-of-sight direction
of the magnetic field (i.e., either toward or away from the Sun) is
either constant for all Zeeman pairs or shows only one organized
reversal in which there exists a line that can be drawn that separates
the side of the SFR where the magnetic field is positive from the side
where it is negative.

\begin{figure}
\begin{center}
\includegraphics[width=5.0in]{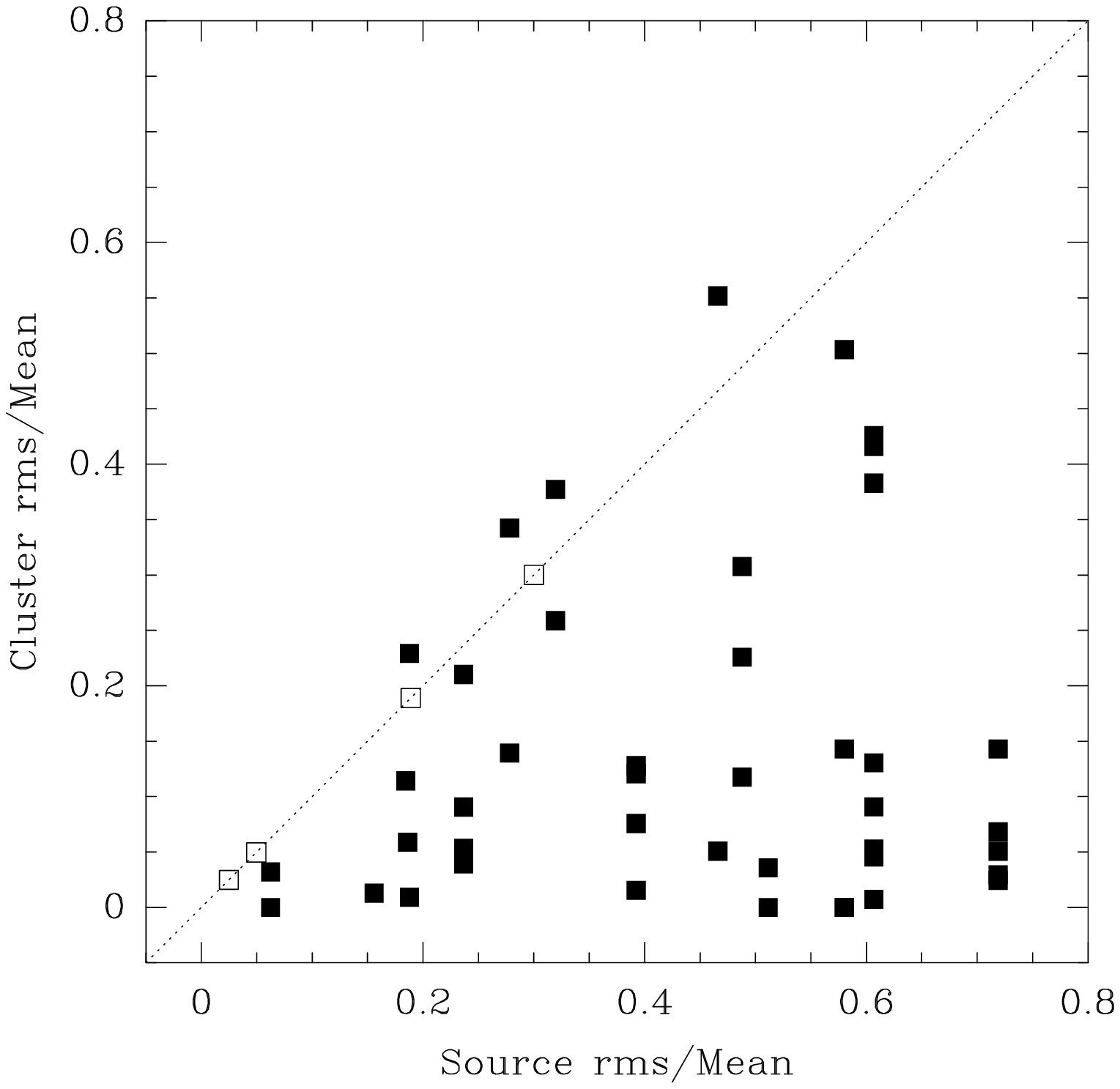}
\end{center}
\singlespace
\caption[Plot of magnetic field strength variation in clusters and
sources.]{Plot of magnetic field strength variation in clusters
(smaller than $\approx 3 \times 10^{15}$~cm) and sources.  The
fractional variation (rms/mean) of the magnetic field strength $|B|$
for each cluster of two or more Zeeman pairs is plotted against the
fractional error for the source as a whole.  G9.622$+$0.195 and
G34.257$+$0.154 are each treated as two separate sources.  Open
squares indicate data points for the middle source in G9.622$+$0.195,
G35.577$-$0.029, G40.622$-$0.137, and S269, in which all Zeeman pairs
are found in a single cluster.  The dotted line indicates equal
fractional errors in cluster and whole-source magnetic field strength.
39 of the 43 filled squares fall below this line, indicating that the
magnetic field variation in a cluster is less than the variation in
the source as a whole.\label{cluster-bsig}}
\doublespace
\end{figure}

The relative consistency of magnetic field strengths in clusters of OH
masers argues in favor of an organized field structure.  When multiple
Zeeman pairs are seen in the same cluster, the range of field
strengths is rarely greater than 2 mG (i.e., $\pm 1$ mG) and often
significantly less.  In no case does the sign of the magnetic field
change between two Zeeman pairs in the same cluster of spots within
$\sim3 \times 10^{15}$~cm.  Figure \ref{cluster-bsig} shows the
fractional variation of magnetic field strength measurements in
clusters compared with the source as a whole.  The intracluster
variation in the magnetic field strength is generally smaller than the
intrasource variation.  If variations in magnetic field strength are
due to variations in density, this suggests that density fluctuations
within a cluster may also be smaller than fluctuations on the scale of
the masing region of a massive SFR.

Given the uniformity of magnetic field direction in the line-of-sight
dimension, it is somewhat surprising that linear polarization vectors
in the same cluster, when converted to magnetic field directions, are
often quite disordered.  Table \ref{sep-ppa} shows the relative
variation of polarization position angle (PPA) as a function of maser
spot separation.  This statistic has a range of [0\degr,90\degr],
since a PPA of angle $x$ is equivalent to one of angle $x + 180\degr$
and the difference cannot exceed $90\degr$.  For pairwise separations
shown in the first column of Table \ref{sep-ppa}, the rms of the
difference in PPA between the two maser spots was calculated.  The
statistic was applied only to spots with a linear polarization
fraction less than 0.707 (equal parts linear and circular for a
totally polarized maser).  This is designed to choose only
$\sigma$-components.
Any statistic comparing the magnetic field direction at
both $\sigma$- and $\pi$-components would have to account
for the natural $90\degr$ difference arising from the PPAs of
$\sigma$- and $\pi$-components in the same magnetic field.  A sample
of maser spots with a uniform random distribution of PPAs would have
an rms of $52\degr$.
The rms value of the PPA \emph{difference} between maser spots is
roughly constant to within the errors for all maser spot separations.
Even at the smallest scale ($< 10^{14}$~cm) the rms in PPA differences
is consistent with a random distribution.  Given the regularity of
magnetic field direction both on source and cluster scales, the PPA
differences cannot be due to magnetic field variations within a
cluster.  Probably Faraday rotation is large enough in most
sources, even on AU scales, to scramble the linear polarization
directions.  See \S \ref{faraday} for further discussion of the
possible effects of Faraday rotation.

\singlespace
\begin{deluxetable}{cccc}
\tablewidth{0pt}
\tablecaption{Variation of PPA with Pairwise Maser Separation\label{sep-ppa}}
\tablehead{
  \colhead{Separation} &
  \colhead{Number of} &
  \colhead{PPA} &
  \colhead{Standard} \\
  \colhead{(cm)} &
  \colhead{Pairs\tnm{a}} &
  \colhead{rms} &
  \colhead{Error\tnm{b}}\\
}
\startdata
0         --- $10^{14}$ & \phn\phn25 & $45\fdg7$ &$9\fdg1$ \\
$10^{14}$ --- $10^{15}$ & \phn189    & $46\fdg6$ &$3\fdg4$ \\
$10^{15}$ --- $10^{16}$ & \phn566    & $49\fdg6$ &$2\fdg1$ \\
$10^{16}$ --- $10^{17}$ & 1707       & $51\fdg8$ &$1\fdg3$
\enddata
\tablewidth{433.62pt}
\tablecomments{Table of rms variations in the PPA between pairs of
maser spots for all sources.  Only pairs of spots whose linear
polarization fraction is less than 0.707 are considered.}
\tnt{a}{Number of pairs of maser spots with linear polarization
fractions less than that shown in the column headings such that the
separation between the maser spots is less than that shown in the
first column.}
\tnt{b}{Standard error of the mean: $\mathrm{rms}/\sqrt{N}$.}
\end{deluxetable}
\doublespace

\singlespace
\subsection{Relation of OH Masers to the \hii\ Region\label{relation-hii}}

The association between OH masers and UC\hii\ regions has been noted
by many authors, including \citet{dieter66} and \citet{mezger67}.  Our
larger sample of OH masers mapped with milliarcsecond resolution
allows us to confirm this finding.  For each continuum source with
nearby masers, we determined an ellipse whose major and minor axes
best matched the overall extent of continuum emission (full width at
zero power for a $4\,\sigma$ detection).  The center of the \hii\
region was taken to be the center of the ellipse, and the radius of
the \hii\ region was taken to be the geometric mean of the semi-major
and semi-minor axes.  The distance of each OH maser from the center of
the \hii\ region, in units of \hii\ region radii, was computed.  Two
possible sources of error are the uncertainty in map registration
between the OH masers and continuum images and the uncertainty in the
assignment of the center of continuum emission in each source.  The
former is estimated to be $0\farcs 3$ ($1\, \sigma$) by \citet{arm}.
The latter may vary depending on source structure.  For a circular
\hii\ region, we estimate a 10\% error, which would correspond to a
$1\,\sigma$ error of $0\farcs 1$ for a typical $1\arcsec$ UC\hii\
region, resulting in a total error of $0\farcs 3$ for the combination
of the two effects.  For large (e.g.,
\object[G5.886-0.393]{G5.886$-$0.393}) or irregularly-shaped (e.g.,
\object[G35.577-0.029]{G35.577$-$0.029}) \hii\ regions, the error in
the estimate of the center of continuum emission may be slightly
higher.

\begin{figure}
\begin{center}
\includegraphics[width=4.0in]{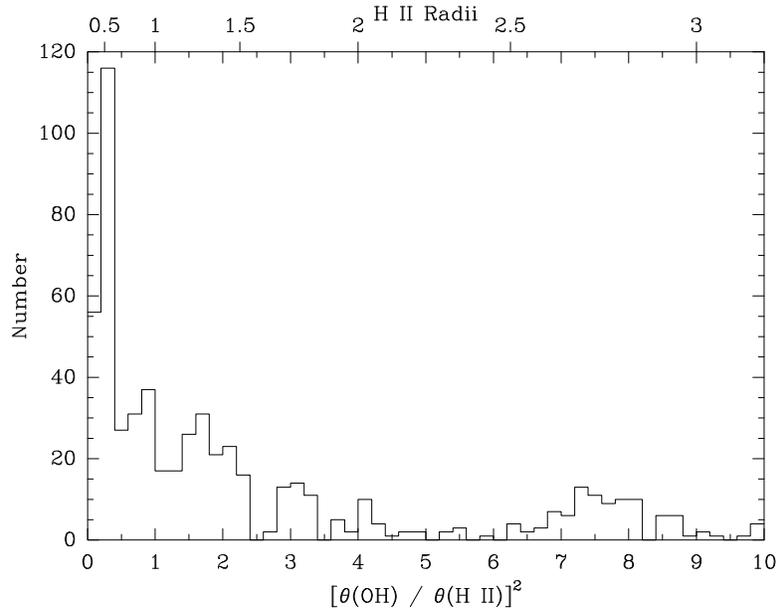}
\end{center}
\singlespace
\caption[Histogram of distances between OH masers and UC\hii\
regions.]{Histogram of distances between OH masers and UC\hii\ regions
shown in units of equal area.  Masers that cannot be identified as
unambiguously associated with a particular \hii\ region have been
excluded; see \S \ref{relation-hii} for details.  We find that 58\% of
OH masers appear within 1.5 radii of the \hii\ regions, suggesting
that OH masers in massive SFRs with \hii\ regions are indeed spatially
associated with them.  The bump at 2.7 \hii\ radii is due primarily to
the masers in ON 2 N; it is probable that these masers are not actually
associated with the \hii\ region shown in Figure 23 of Paper I.  A
small tail of the distribution out to 6 \hii\ radii has been
suppressed for clarity.\label{mas-hii}}
\doublespace
\end{figure}

The distribution of the distance of OH masers, normalized by area,
from the center of the \hii\ region is shown in Figure \ref{mas-hii}.
Masers in \object[G9.622+0.195]{G9.622$+$0.195},
\object[G34.257+0.154]{G34.257$+$0.154}, and \object{Cep A} were
treated as containing three, two, and four continuum sources
respectively, and masers distances were calculated from the nearest
source.  Several sources were excluded from this analysis because no
nearby continuum source was detected: the northern grouping of
G9.622$+$0.195, G40.622$-$0.137, \object[G196.454-01.677]{S269},
\object{Mon R2}, and \object[G351.775-0.538]{G351.775$-$0.538}.
Additionally, W75 N and the two southeastern continuum sources of Cep
A were excluded since it is not always clear which of several
continuum sources to associate a maser spot with.

Including \object[Onsala 2N]{ON~2~N} and G40.622$-$0.137, 50\% of maser spots are located
within 1.5 radii of a UC\hii\ region; this number grows to 58\% when
these two sources are excluded.  The distribution of maser
spots in these two sources is clearly offset from the \hii\ region,
suggesting that the masers may be associated with a second, undetected
continuum source nearby.  The large peak near 0.5 \hii\ region radii
in Figure \ref{mas-hii} is due mainly to the western cluster of maser
spots in G5.886$-$0.393.
Overall, it appears that the distribution of OH masers peaks near the
center of the \hii\ region, consistent with \citet{garay}.
The tail of the distribution of maser spots at several
radii from the UC\hii\ region may represent spots associated with
another star, not with the nearest detectable continuum source, as in
ON~2~N and W51 (e1 and e2).

These results are not consistent with a random distribution of OH
maser spots within a shell around the \hii\ region.  A uniform random
distribution would peak at a projected radius $r_\mathrm{proj} > 1$
(in \hii\ region radii).  For density distributions falling off as
$r^{-1}$ or $r^{-2}$, the peak of the distribution will be at
$r_\mathrm{proj} = 1$.  In all cases, the distribution of OH masers
would be expected to double across $r_\mathrm{proj} = 1$ because
masers located behind the \hii\ region would be obscured by the
UC\hii\ region, which is optically thick at $\lambda = 18$~cm.  We see
no evidence for this discontinuity at 1 \hii\ region radius in our
data.

\begin{figure}
\begin{center}
\includegraphics[width=5.0in]{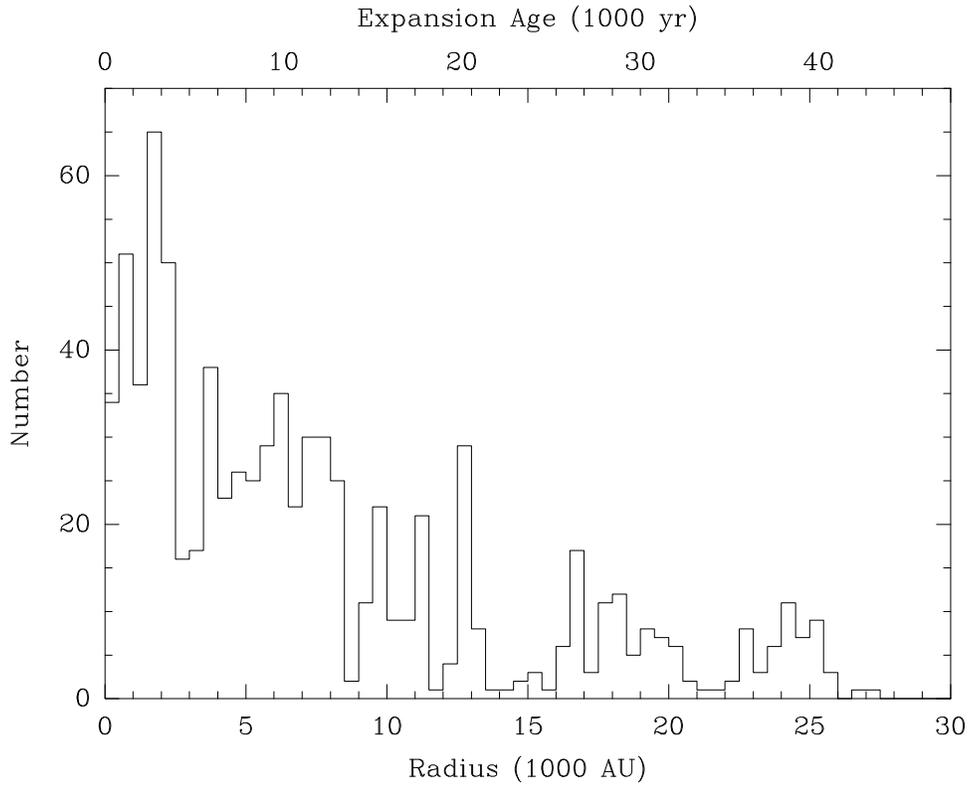}
\end{center}
\singlespace
\caption[Histogram of absolute distances between OH masers and UC\hii\
regions.]{Histogram of absolute distances between OH masers and
UC\hii\ regions.  The data are as in Figure \ref{mas-hii} but plotted
in distance units and not normalized by area.  The expansion age,
defined as radius divided by expansion speed, shown at the top would
be appropriate for expansion at 3 \kms, as measured for 
W3(OH).\label{mas-nohii}}
\doublespace
\end{figure}

Figure \ref{mas-nohii} shows the distribution of projected linear
distances of OH masers from the center of the associated UC\hii\
region.  82\% of maser spots are located within 13000 AU of the center
of the \hii\ region.  Also shown is the assumed dynamical age of the
masers for an expansion speed of 3 \kms.  This is consistent with the
speed measured in W3(OH) from both proper motion of OH masers
\citep{w3oh} and direct expansion of the UC\hii\ itself
\citep{kawamura}.  The dynamical age of most masers is less than $2
\times 10^4$ yr.

\begin{figure}
\begin{center}
\includegraphics[width=5.0in]{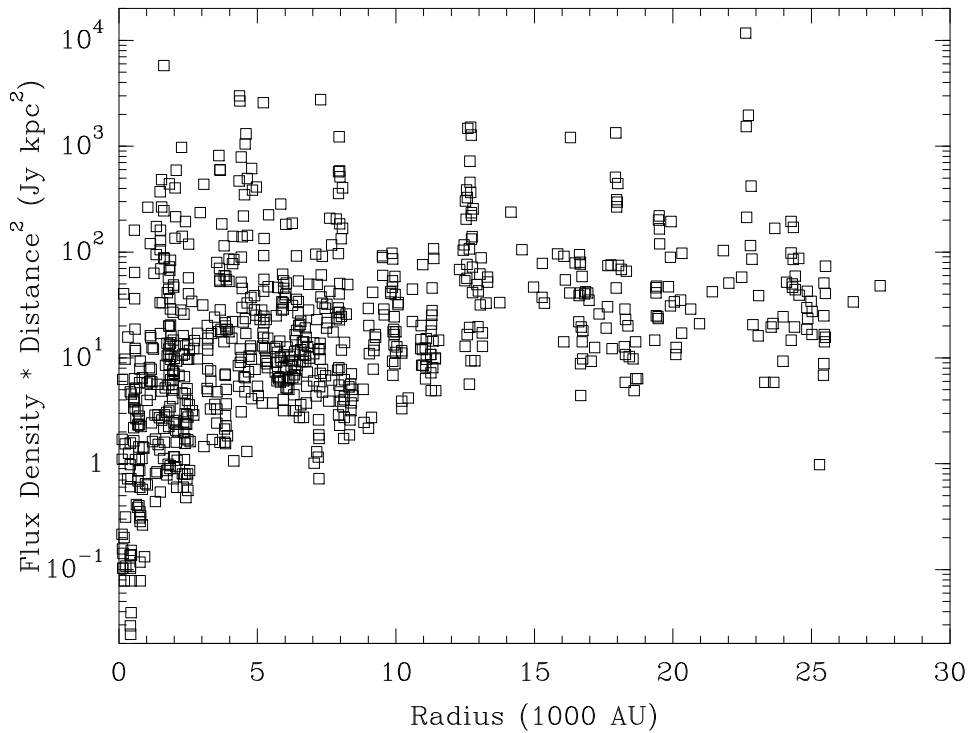}
\end{center}
\singlespace
\caption[Distribution of maser power per bandwidth with
radius.]{Distribution of maser power per bandwidth with radius.  All
flux densities are multiplied by the square of the distance to the
source in kiloparsecs.  The horizontal axis shows the distance of each
maser spot from the center of the associated UC\hii\ region.  Multiple
\hii\ regions in the same source (e.g., G9.622$+$0.195) are considered
independently.  Data are not plotted when no nearby UC\hii\ region is
seen (e.g., S269 and G351.775$-$0.538) or when it is unclear which
\hii\ region to match maser spots with (W75 N).  There does not appear
to be a correlation between the power per bandwidth and the distance
of maser spots from the UC\hii\ region.  The slight dip near zero
radius is due to the inclusion of Cep A, whose proximity allowed
detection of spots of weaker normalized flux density than for other
sources.\label{flux-rad}}
\doublespace
\end{figure}

\begin{figure}
\begin{center}
\includegraphics[width=5.0in]{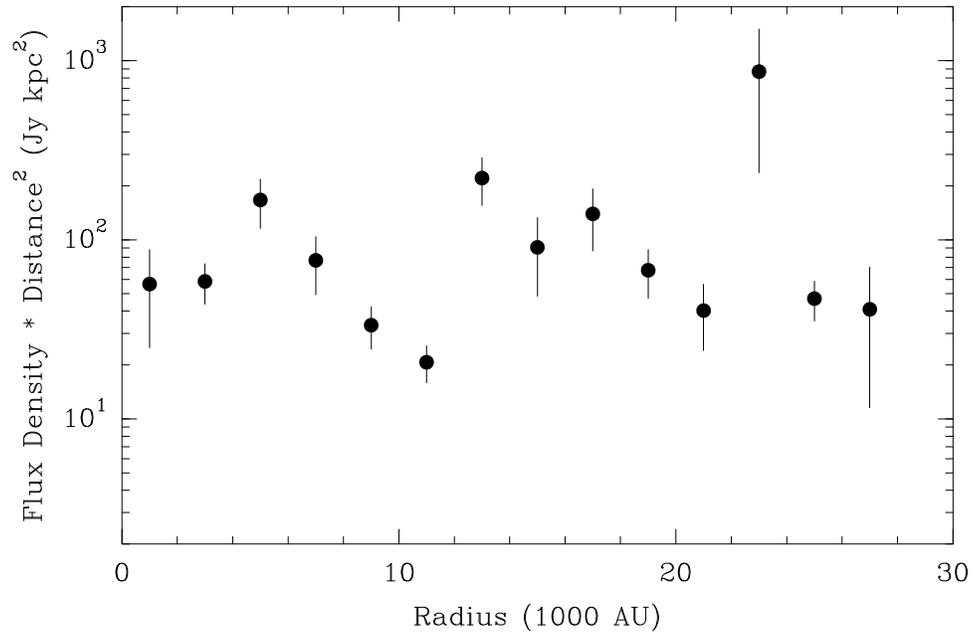}
\end{center}
\singlespace
\caption[Distribution of mean maser power per bandwidth with
radius.]{Distribution of mean maser power per bandwidth with radius.
The plot shows the mean of the maser flux normalized to a distance of
1~kpc as well as the standard error of the mean
($\mathrm{rms}/\sqrt{N}$).  Errorbars are not symmetric because a
linear average is plotted on a logarithmic scale.  The data are binned
by units of 2000 AU.  The mean power per bandwidth and the distance of
maser spots from the UC\hii\ region is constant with distance,
although large deviations are possible due to source-to-source
differences.  See Figure \ref{flux-rad} for more details.
\label{meanflux-rad}}
\doublespace
\end{figure}

\begin{figure}
\begin{center}
\includegraphics[width=5.0in]{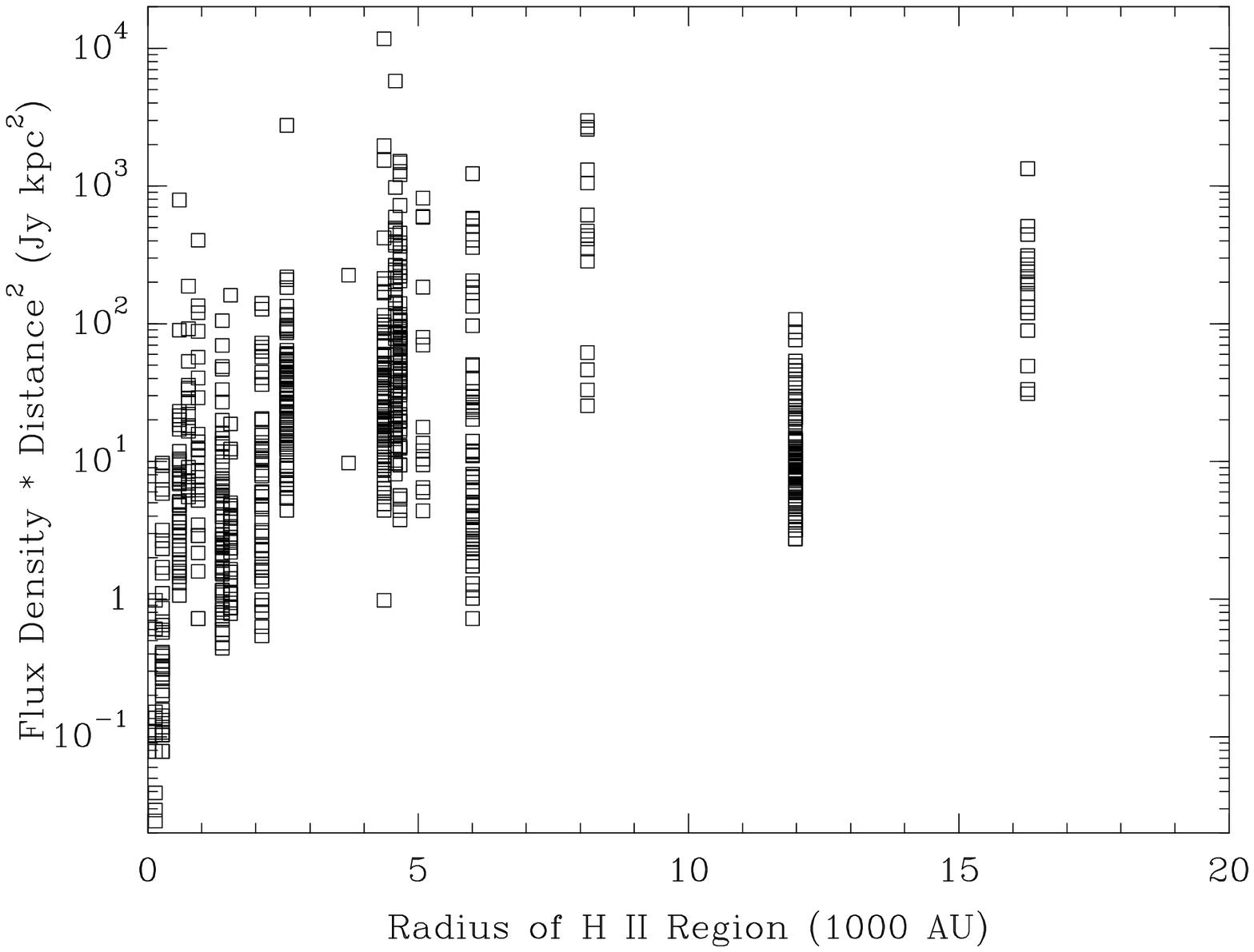}
\end{center}
\singlespace
\caption[Distribution of maser power per bandwidth with \hii\ region
size.]{Distribution of maser power per bandwidth with \hii\ region
size.  The horizontal axis shows the geometric mean of the semi-major
and semi-minor axes of the UC\hii\ region, converted to linear
distance.  See Figure \ref{flux-rad} for more details.  There does not
appear to be a correlation between the size of the \hii\ region and
the power per bandwidth of the maser spots surrounding it.  Since
the size of an \hii\ region is a measure of its age, this suggests
that OH masers do not become systemically fainter over the range of
ages of the \hii\ regions in our sample.  As in Figure \ref{flux-rad},
the dip near zero radius is due to Cep A.\label{flux-hii}}
\doublespace
\end{figure}

While the number of OH maser spots cuts off at a dynamical age of
several $\times 10^4$ yr, individual OH masers do not appear to fade
appreciably during this time.  Figure \ref{flux-rad} shows the
relation of maser power per bandwidth (i.e., flux density normalized
to a constant distance) to the separation between the UC\hii\ region
center and the masers.  Figure \ref{meanflux-rad} shows the mean power
per bandwidth and standard error of the mean for the same data.  The
maser power per bandwidth appears to be constant with distance from
the center of the \hii\ region.  If the distance of the masers from
the \hii\ region is indeed correlated with their age, OH masers do not
become systemically brighter (or fainter) with age, at least not in
the $4 \times 10^4$ yr timescale our data span.  Figure \ref{flux-hii}
shows the relation of maser power per bandwidth to the size of the
associated UC\hii\ region.  Since \hii\ regions undergo expansion,
their size is a measure of the age of the system.  Again, there does
not appear to be a correlation between the maser power per bandwidth
and the age of the system.  However, we do not see any maser spots
located more than 30000 AU (0.15 pc) from the center of the associated
UC\hii\ region (including G351.775$-$0.538, for which the nearest
UC\hii\ region is several arcseconds away).

\citet{habing79} have observed that OH masers are not seen around
\hii\ regions once they leave the ultracompact phase ($d < 0.15$ pc).
Indeed, not only are OH masers not seen around ``compact \hii\ (C\hii)
regions'' ($0.1 < d < 1$~pc), they are not seen at comparable radii
around \emph{ultra}compact \hii\ regions.  The lack of OH masers at
large distances from the associated ultracompact \hii\ region was
first noted by \citet{habing74}, who suggested that OH maser phenomena
disappear at a radius of 15000 AU (0.07 pc).  Our larger sample size
at much higher angular resolution indicates that there is a sharp
cutoff at about twice this radius.  It is possible that the physical
conditions (such as temperature and density) responsible for maser
activity do not exist at large radii.  Alternatively, the ionization
front catches up to the shock front as the \hii\ region expands into
an environment whose density decreases with radius, thereby destroying
the OH masers, which are believed to exist in the region between the
ionization and shock fronts (see \S \ref{shocks}).

\subsection{Relation of OH Masers to the Surrounding Material}

In order to interpret the bulk motions of OH masers, it is necessary
to determine the velocity of the massive star or stars being formed.
Frequently this is determined through hydrogen recombination line
observations, which give information on the velocity of the ionized
\hii\ region surrounding the central star.  But recombination lines
are not well suited to the task.  Recombination lines are subject to
sometimes severe Doppler and impact broadening.  Even at high
frequency, different recombination lines can be biased by a few
\kms~\citep{berulis83,sams96} due to differing optical depths in an
expanding \hii\ region.

We have observed the $(J,K) = (1,1)$ line of ammonia in order to
provide a context for the OH maser observations.  \citet{rmb} argue,
based on the similarity of distribution of NH$_3$ absorption and OH
maser emission in W3(OH), that the two species are found in the same
clumps of material.  The physical conditions they deduce from NH$_3$
$(1,1)$ and $(2,2)$ observations are consistent with the physical
conditions necessary for OH maser activity.  Also, the velocity of
peak absorption in NH$_3$ closely matches the average velocity of the
OH masers.  Unfortunately, in many sources NH$_3$ absorption is not
clearly detected, and NH$_3$ emission velocities must be used instead.
NH$_3$ \emph{emission} is a less reliable indicator of the velocity of
the central star, since emission usually traces motions on a larger
scale (often $> 1\arcmin$).  In the region of W3(OH), ammonia emission
velocities differ from the average OH maser velocity by approximately
4~\kms~\citep{wilson93}.  It is possible that systemic biases of a few
\kms\ are introduced using emission velocities, but there is no way to
obtain the radial velocity of the star to greater accuracy.

The spectra in Figures 40 to 42 of Paper I are provided for regions of
emission or absorption located roughly coincident with the extent of
OH masers on the sky.  It is important to remember that while OH
masers exist primarily near a UC\hii\ region, the neutral NH$_3$ may
exist at a wide range of radii.  Thus the inferred NH$_3$ velocities
could in principle be affected by motion of material quite distant
from the \hii\ region.  The NH$_3$ velocity may be shifted from the
rest velocity of the star by an amount comparable to the velocity
dispersion of the molecular cloud.  From the virial theorem, the
velocity dispersion is
\begin{equation}
\Delta v = \sqrt{\frac{GM}{2R}},
\end{equation}
where $G$ is the gravitational constant, $M$ is the total mass, and
$R$ is the radius of the sphere.  For a cloud with $M = 100 M_\sun$ and
a radius of 0.1 pc, $\Delta v \approx 1.5$~\kms.

When the velocity of the nearby NH$_3$ is measured, it generally falls
toward the middle of the range of OH maser velocities, as shown in
Figures 40 to 42 of Paper I.  There are some minor exceptions to this
rule.  In G35.577$-$0.029 and Mon R2, the NH$_3$ velocity is near an
extremum of OH maser velocities.  (In \object[Onsala 1]{ON~1} the OH
masers fall into two disjoint groups at $2.5 - 6$ \kms\ and $13 - 17$
\kms, and NH$_3$ emission detected in a broad region located 5\arcsec\
to 30\arcsec\ north of ON~1 (not shown in the panel) falls near
11~\kms.)

\begin{figure}
\begin{center}
\includegraphics[width=5.0in]{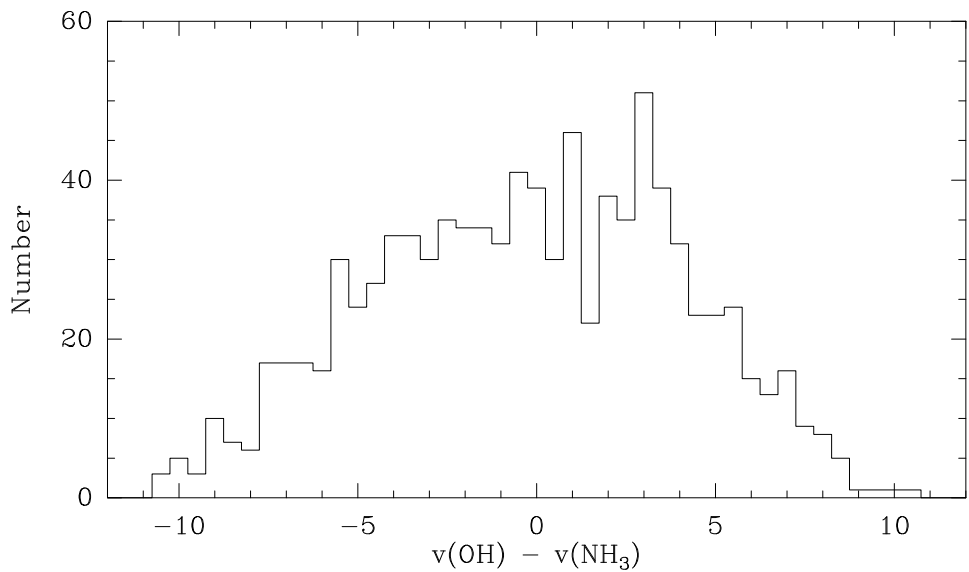}
\end{center}
\singlespace
\caption[Histogram of differences between OH and NH$_3$
velocities.]{Histogram of differences between OH and NH$_3$ velocities
for those sources in Figures 40 to 42 of Paper I with an NH$_3$
velocity indicated.  The median velocity difference is $-0.30$~\kms\
with a standard deviation of 1.89 \kms.  The mean is $-0.29$~\kms\
with a standard error of the mean (rms/$\sqrt{N}$) of 0.06~\kms,
assuming all masers have independent velocities.  Taking clumping into
account, the standard error of the mean could be higher by a factor of
$\approx 3$.
\label{vel-nh3}}
\doublespace
\end{figure}

\begin{figure}
\begin{center}
\includegraphics[width=5.0in]{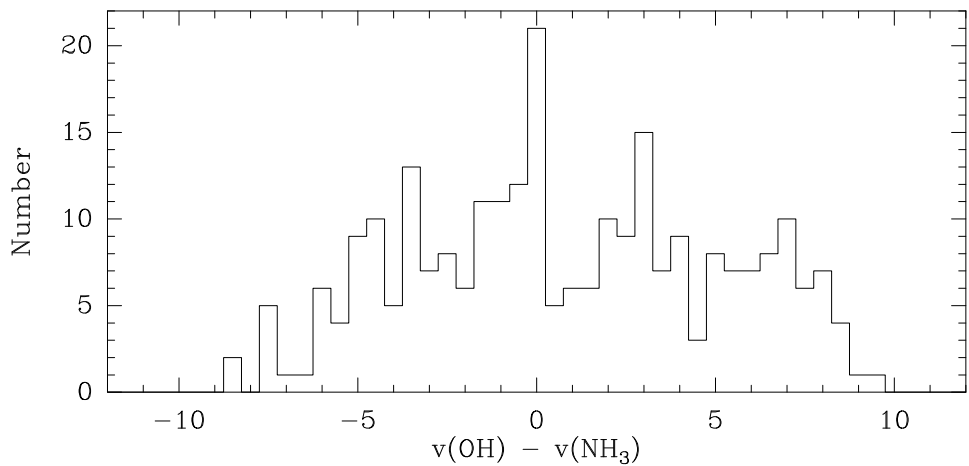}
\end{center}
\singlespace
\caption[Histogram of differences between OH and NH$_3$ velocities
atop \hii\ regions.]{Histogram of differences between OH and NH$_3$
velocities for those maser spots located within one radius of the
associated UC\hii\ region.  The median velocity difference is
$+0.22$~\kms\ with a standard deviation of 1.90~\kms.  The mean is
$0.79$~\kms\ with a standard error of the mean of 0.12~\kms.
\label{vel-nh3-atop}}
\doublespace
\end{figure}

Figure \ref{vel-nh3} shows a histogram of the differences between OH
maser velocities and the adopted NH$_3$ velocity for all
sources having detected NH$_3$.  Of the 926 OH maser spots, 51.7\% are
blueshifted with respect to the NH$_3$ velocity, and 48.3\% are
redshifted.  The median velocity difference is $-0.30$~\kms\ with an
rms of $1.89$~\kms, while half the differences fall within the range
$-3.52$ to $+3.01$~\kms.  Since a zero difference falls comfortably
within this range, we cannot confidently state that OH masers are
consistently blueshifted or redshifted with respect to the surrounding
material, as might be expected if a single type of motion, such as
expansion or contraction, dominates OH maser kinematics.  When only
maser spots located within one projected \hii\ region radius are
considered ($N = 261$), the median velocity difference is
$+0.22$~\kms\ with an rms of $1.90$~\kms\ (Figure \ref{vel-nh3-atop}),
with half the differences falling in the range $-2.58$ to
$+3.98$~\kms.
In either case, there does not appear to be a detectable difference
between the OH and NH$_3$ velocities to within our errors.
If expansion dominates
the dynamics of the masing regions, the masers projected atop the
UC\hii\ region (and therefore in front of it, since UC\hii\ regions
are in general optically thick at $\lambda = 18$ cm) should be
blueshifted with respect to the large-scale ambient material.
However, the opposite appeared to be true of W3(OH) \citep{reidw3},
although later proper motion measurements of the OH masers
definitively established that they are expanding \citep{w3oh}.  It is
worth pointing out however that in G43.796$-$0.127, where nearly all
of the masers are projected against the UC\hii\ region, the OH masers
are preferentially blueshifted with respect to the NH$_3$ emission,
mildly suggestive of expansion.

Because of the aforementioned possibility of systemic errors of a few
\kms\ in determining the radial velocity of the central star from
NH$_3$ velocity measurements, we cannot identify whether a single type
of motion, such as gravitational infall or slow expansion, dominates
the kinematics of OH masers in massive star-forming regions.  We can
in general rule out kinematic modes in which the OH masers would be
moving at tens of \kms\ or more, such as a freely expanding \hii\
region at $\gtrsim 10$~\kms.

\section{Discussion\label{discussion}}

\subsection{Saturation\label{saturation}}

Interstellar OH masers are most likely saturated \citep{reidw3}.  The
saturation temperature, $T_s$, for OH masers is given by
\begin{equation}
T_s = \frac{h\nu}{2k}\frac{\Gamma}{A}\frac{4\pi}{\Omega}
\end{equation}
\citep{reidmoranbook}, where $A$ is the Einstein coefficient, $\Gamma$
is the decay rate, and $\Omega$ is the solid angle of beaming.  For
$\Gamma$ = 0.03~s$^{-1}$ as typical for a far-infrared rotational
transition likely pumping the maser, the saturation temperature is $(2
\times 10^{8})\, \Omega^{-1}$~K.  The most compact maser component
from the space-VLBI observation of G34.257$+$0.154 by
\citet{slyshspace} has a beaming angle $\Omega \approx 0.01$, which
corresponds to a maximum saturation temperature $T_s \approx 2 \times
10^{10}$~K.
For the most part, the brightness temperatures ($T_B$) listed in
Tables 2 through 20 of Paper I are below this value.  But these are
apparent brightness temperatures calculated from the undeconvolved
spot size, which are likely scatter broadened for most sources.

Many of our spots appear to be partially resolved, as shown in Table
\ref{deconv-table}.  However, several caveats apply to the deconvolved
spot parameters.  First, the measured spot size may be larger than the
physical spot size due to interstellar scattering.  Second, large
deconvolved spot sizes may be the result of misidentifying spatially
blended maser spots as a single spot.  Third, determining the
deconvolved spot size of a small maser spot is less accurate than for
a large maser spot, because the deconvolved spot size is obtained from
differencing two larger numbers (the squares of the undeconvolved spot
size and the beam size).
The net of all three effects is that maser spot sizes are probably
smaller than that calculated from deconvolution, and that the
overestimation may be greater for heavily scatter-broadened sources.
Even a deconvolved spot size would underestimate the actual $T_B$ by
the square of the ratio of the apparent spot size to the unbroadened
spot size.

For a typical FWHM spot size of 3~mas in W75~N (which has
very little scatter broadening), $T_s = 5 \times 10^{9}$~K.  This
corresponds to a flux density of $S_\nu = 2 k T_s \Omega_\mathrm{spot}
\lambda^{-2}$, where $\Omega_\mathrm{spot}$ is the solid angle
subtended by the maser spot.  Taking $\Omega_\mathrm{spot} \leq
1.7 \times 10^{-16}$~sr, the saturation value $S_\nu \leq 7 \times
10^{-25}$~erg~cm$^{-2}$~s$^{-1}$~Hz$^{-1}$~sr$^{-1}$, or 0.07~Jy.
This is near our detection limit, so nearly all spots that we detect
are partially saturated if the spot sizes for W75~N are typical.

\begin{deluxetable}{lccccc}
\tablewidth{0pt}
\tablecaption{Deconvolved Spot Sizes\label{deconv-table}}
\tablehead{
  \colhead{} &
  \colhead{Distance} &
  \colhead{Number of} &
  \colhead{Median Deconvolved} &
  \multicolumn{2}{c}{90\% Range of Spot Sizes} \\
  \colhead{Source} &
  \colhead{(kpc)} &
  \colhead{Spots} &
  \colhead{Spot Size (mas)} &
  \colhead{Min (mas)} &
  \colhead{Max (mas)} \\
}
\startdata
G5.886$-$0.393   & \phn3.8 & \phn98 &    19.66 &    12.44 &    30.33 \\
G9.622$+$0.195   & \phn5.7 & \phn38 &    16.44 &    10.86 &    21.05 \\
G10.624$-$0.385  & \phn4.8 & \phn14 &    23.55 &    19.48 &    28.96 \\
G34.257$+$0.154  & \phn3.8 & \phn88 & \phn5.46 & \phn0.00 &    13.50 \\
G35.577$-$0.029  &    10.5 & \phn15 &    16.22 &    12.39 &    19.55 \\
G40.622$-$0.137  & \phn2.2 & \phn39 &    19.68 &    15.57 &    24.22 \\
G43.796$-$0.127  & \phn9.0 & \phn60 & \phn7.96 & \phn0.00 &    13.11 \\
W51 e1           & \phn7.0 & \phn97 &    10.51 & \phn6.62 &    20.07 \\
W51 e2           & \phn7.0 & \phn94 & \phn9.53 & \phn5.19 &    15.23 \\
ON 1             & \phn3.0 & \phn49 & \phn2.68 & \phn0.00 & \phn7.30 \\
K3$-$50          & \phn8.7 & \phn17 &    31.46 &    23.45 &    34.82 \\
ON 2 N           & \phn5.6 & \phn73 &    25.29 &    13.69 &    37.11 \\
W75 S            & \phn2.0 & \phn65 & \phn4.17 & \phn1.57 & \phn7.75 \\
W75 N            & \phn2.0 &    120 & \phn3.14 & \phn0.00 & \phn8.90 \\
Cep A            & \phn0.7 & \phn62 & \phn6.77 & \phn1.95 &    12.96 \\
NGC 7538         & \phn2.8 & \phn30 & \phn6.83 & \phn3.19 &    11.20 \\
S269             & \phn3.8 & \phn19 & \phn6.08 & \phn1.23 & \phn9.47 \\
Mon R2           & \phn0.9 & \phn27 & \phn7.86 & \phn2.20 &    13.49 \\
G351.775$-$0.538 & \phn2.2 & \phn50 &    67.40 &    45.35 &    76.07
\enddata
\tablecomments{Spot sizes may be overestimates.  See \S
  \ref{saturation} for details.}

\end{deluxetable}

\subsection{Faraday Rotation\label{faraday}}

Faraday rotation can complicate the interpretation of linear
polarization in two ways.  First, external Faraday rotation between a
maser and the observer will cause the polarization position angle
(PPA) of linear polarization to rotate, making interpretation of the
magnetic field direction on the plane of the sky more difficult.
Second, internal Faraday rotation along the amplification path may
decrease the linear polarization fraction of the radiation, completely
circularizing it if the Faraday rotation is strong enough
\citep{gkk2}.  Since this also reduces the effective gain length for
linear polarization, Faraday rotation may also prevent otherwise
highly linearly-polarized maser components from being amplified to the
limits of detectability.  Thus, spots with a large linear polarization
fraction (e.g., $\pi$-components and $\sigma$-components where the
magnetic field is near the plane of the sky) may be suppressed
relative to spots with a small linear polarization fraction (e.g.,
$\sigma$-components where the magnetic field is directed along the
line of sight).

External Faraday rotation in the interstellar medium between a maser
and the observer would cause a rotation of the PPA of the linear
polarization of each spot, given by
\begin{equation}
\mathrm{RM} = 8.1 \times 10^{5} \int n_e B_\parallel \, dl,
\end{equation}
where RM is the rotation measure in rad m$^{-2}$, $n_e$ in cm$^{-3}$,
$B_\parallel$ is the component of the magnetic field parallel to the
direction of propagation in G, and $dl$ in is the differential path
length along the line of sight in pc \citep{thompson}.  In some
regions, such as the northern cluster in W75 N, the linear
polarization vectors are predominantly aligned along the line of maser
spots (see \S \ref{triplet}).  An RM of about 10 rad m$^{-2}$ would
produce a rotation of the PPAs in a source of $20\degr$.
A rotation of the polarization vectors by an amount greater than
this would cause the vectors to no longer appear to be aligned with
larger structures, unless the rotation was near a multiple of $180\degr$.
According to the ATNF Pulsar Catalogue
\citep{manchester05}\footnote{The catalogue is available online at
http://www.atnf.csiro.au/research/pulsar/psrcat .}, the only pulsar
with known rotation measure located within $10\degr$ of W75 N at
comparable heliocentric distance is B2021+51, for which the RM is
$-6.5$~rad~m$^{-2}$ \citep{manchester72}.

Internal Faraday rotation over the region of amplification may destroy
linear polarization in both $\sigma$- and $\pi$-components, possibly
suppressing $\pi$-components altogether.
The Faraday rotation over a region with average electron
density $n_e$ and parallel magnetic field strength $B_\parallel$ is
\begin{equation}
\psi = 0 \fdg 05 \left(\frac{n_e}{1~\mathrm{cm}^{-3}}\right)
                \left(\frac{B_\parallel}{1~\mathrm{mG}}\right)
                \left(\frac{L}{10^{14}~\mathrm{cm}}\right)
                \left(\frac{\lambda}{18~\mathrm{cm}}\right)^2,
\end{equation}
where $\lambda$ is the wavelength of the transition.  For a typical
ground-state ($\lambda = 18$~cm) OH maser, $B \approx 5$~mG.  The
effective amplification length $L$ is likely to be less than the
clustering scale due to velocity coherence.  A crude
estimate is that $L \approx D \Delta v/\Delta V$, where $D =
10^{15}$~cm is the diameter of the masing cloud, $\Delta v =
0.2$~\kms\ is a typical maser line width, and $\Delta V \approx
2$~\kms\ is a reasonable velocity shift across the cloud based on
observations of W3(OH) \citep{reidw3} and theoretical modelling
\citep{pavlakis96}.  Thus, for an effective amplification length $L =
10^{14}$~cm, an electron density of about 300~cm$^{-3}$ would be
sufficient to produce a rotation of $90\degr$ along the path of
amplification.  For H$_2$ densities of $10^5$ to $10^8$ as is typical
in OH masing regions \citep{cragg}, this would require a fractional
ionization ($n_e / n_{\mathrm H_2}$) of $3 \times 10^{-6}$ to $3
\times 10^{-3}$.  This is higher than the ionization rate that would
be expected from cosmic-ray ionization alone \citep{shubook}, but
consistent with the $10^{-4}$ that occurs in the \ion{C}{2} regions
around \hii\ regions where OH masers may exist \citep{sternberg,
garcia}.  Ionized carbon and to a lesser extent sulfur may play an
important role in producing free electrons, due to their abundance and
ease of photoionization.  While the hydrogen in the \hii\ region
absorbs all the ultraviolet photons with energies greater than
13.6~eV, many softer photons pass through undisturbed.
\citet{sternberg} calculate that the ionization fraction may be
slightly greater than $10^{-4}$ in the \ion{C}{2} region, located
around the \hii\ region, and about $10^{-5}$ in the \ion{S}{2} region,
in turn located around the \ion{C}{2} region.  Based on their models
of photon-dominated regions as well as the locations of OH maser spots
just outside \hii\ regions, it is likely that they exist near or
embedded in the \ion{C}{2} regions.

We can form a consistent picture of linear polarization in OH masers
if the amount of Faraday rotation in a typical maser source is near a
critical point, i.e., such that the product $n_e B_\parallel L
\approx$ several $\times 10^{17}$~cm$^{-2}$~mG.  Based on maser line
widths and brightness temperatures, the amplification length is
typically at least 20 unsaturated gain lengths and probably greater
for highly saturated masers \citep{reidmoranbook}.  A typical maser
spot has a significant amount ($> 1$~rad) of Faraday rotation over the
amplification length but a small amount ($< 1$~rad) over a single gain
length.  In this case, some linear polarization will survive
amplification, but Faraday rotation scrambles the PPA of the linear
polarization, so it will not be simply interpretable as a magnetic
field direction.  If the Faraday rotation is a factor of $\sim 5$
smaller, the Faraday rotation over the amplification length will be
small, so high linear polarization fractions may be observed, and the
PPA may still be correlated with the magnetic field direction.  On the
other hand, if the Faraday rotation is a factor of $\sim 5$ larger,
the Faraday rotation over a gain length can be large, and linear
polarization fractions will approach zero \citep{gkk2}.  In cases
where the Faraday rotation per gain length is significant,
interpretation of the PPA will be difficult because maser
amplification will stimulate emission in the orthogonal linear mode as
well \citep{melrose04}.  Significant generalized Faraday rotation may
also circularize $\pi$-components.

An example of a source with small internal Faraday rotation over the
entire amplification length is W75 N, in which $\pi$-components are
detected in abundance, especially in the northernmost group of maser
spots (see \S \ref{triplet}).  As Figure \ref{w75n-ppa} shows, $\pi$-
and $\sigma$-components are easily identifiable in this group based on
the PPA of the linear polarization.  It is interesting to note that
modelling of the OH masers in W75 N by \citet{gray03} indicates that
the maser amplification length is several orders of magnitude smaller
than that typically assumed in other sources, although the density is
also higher.  For the range of ionization fractions given above, the
resulting Faraday rotation would be less than 1 radian over the
amplification length.

At the other extreme is W51 e1 and e2, in which practically no linear
polarization whatsoever is detected.  This is consistent with
significant Faraday rotation along a gain length, which would suppress
the amplification of $\pi$-components and circularize the otherwise
elliptically-polarized $\sigma$-components.  Such Faraday
depolarization may also explain certain maser features that are seen
with similar flux densities in RCP and LCP but without any detected
linear polarization, such as spots 18, 24, and 36 in Table 9 of Paper
I.  It is highly unlikely that the lack of linear polarization in W51
can be due to chance alignments of the magnetic field in an extremely
narrow cone oriented toward or away from us at each maser site,
producing $\sigma$-components that are purely circularly polarized,
because the magnetic field is seen to reverse line-of-sight direction
across the source.  Thus, the inclination of the magnetic field to the
line of sight must take on values intermediate to the $0\degr$ and
$180\degr$ required for pure-circular masers in the absence of Faraday
rotation.

A medium range of Faraday rotation would be enough to partially (but not
totally) circularize $\sigma$-components.  The observed circular
polarization fraction will in general be a complicated function of the
electron density, maser gain length, and angle of propagation with
respect to the magnetic field direction, but the presence of Faraday
rotation will strictly increase the circular polarization fraction
compared to the case in which no Faraday rotation is present
\citep{field94}.  Straightforward application of equation (50) of
\citet{gkk2} without accounting for Faraday rotation will cause the
inclination of the magnetic field to the line of sight to be
underestimated.

Faraday rotation may also explain why linear polarization vectors
appear to be disorganized in some sources.  If a large electron
density is required for Faraday rotation along the amplification path
in the interior of a masing cloud, it is likely that the electron
density is high exterior to the masing cloud as well.  This would
rotate the apparent PPA of the maser emission.  Fluctuations in the
electron density, possibly caused by density inhomogeneities or
anisotropy of the ionizing radiation field, could cause emission from
adjacent maser spots to be Faraday rotated by different amounts.  If
this is indeed the case, reconstructing the magnetic field orientation
in the plane of the sky is a difficult task, and reconstructing the
full three-dimensional orientation of the magnetic field may be nearly
impossible.

\begin{figure}
\begin{center}
\includegraphics[width=5.0in]{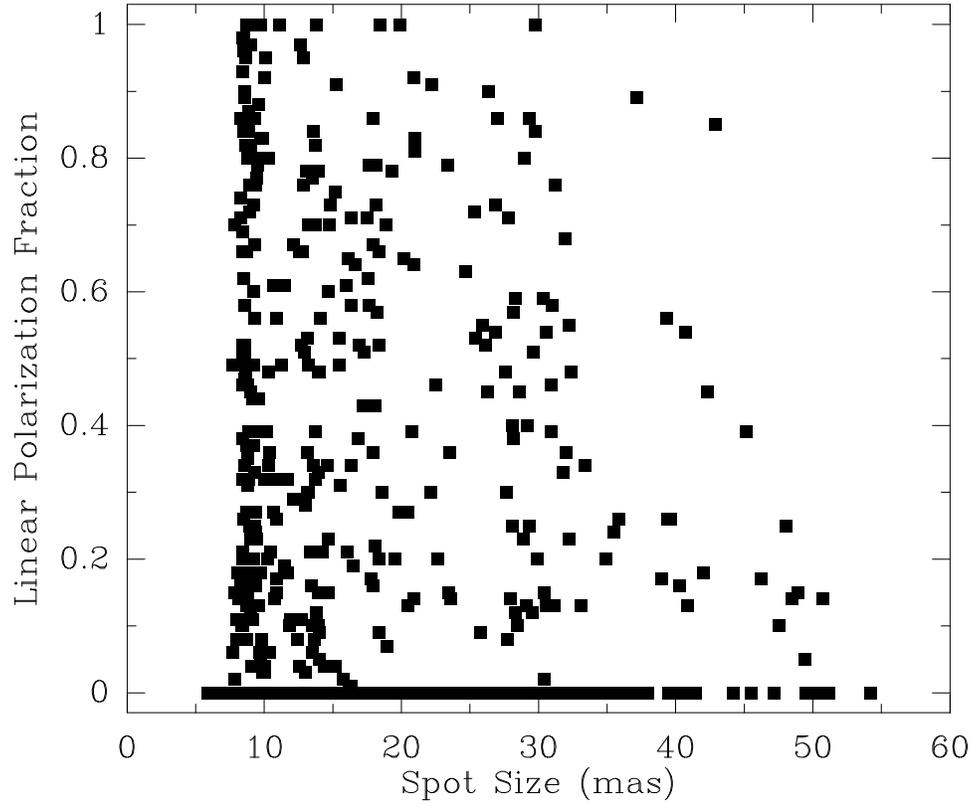}
\end{center}
\singlespace
\caption[Linear polarization fraction as a function of spot
size.]{Plot of linear polarization fraction as a function of spot
size.  The spot size is taken to be the geometric mean of the
undeconvolved spot fit major and minor axes.  The relative lack of
spots with a high linear polarization fraction at large spot sizes may
be due to blending of adjacent spots within the beam.  The linear
polarization fraction does not appear to be otherwise correlated with
the observed spot size.\label{ml-spot}}
\doublespace
\end{figure}

Scatter broadening of some sources implies strong density fluctuations
and a high column density of electrons along the radiation propagation
path.  Since the scatter broadening is proportional to the distance
between the source and the scattering screen \citep[e.g.,][]{boyd72},
a screen of electrons near the source is unlikely to increase the
angular size as much as a cloud of electrons several kiloparsecs away
in the Galactic plane.  The lack of correlation between scattered size
and Faraday depolarization provides further evidence that the
scattering is external to the masing regions.  W51 is not a
particularly scatter-broadened source, yet essentially no linear
polarization is detected.  G351.778$-$0.538 is heavily
scatter-broadened, but several spots with a high degree of linear
polarization are seen.  Figure \ref{ml-spot} shows a plot of the
linear polarization fraction as a function of spot size for all
sources combined.  There does not appear to be a correlation of the
linear fraction with the size of the observed maser spots.  This
suggests that the electron screen responsible for scattering is
Galactic in origin.

\subsection{Total Polarization\label{totpol}}

\begin{figure}
\begin{center}
\includegraphics[width=5.0in]{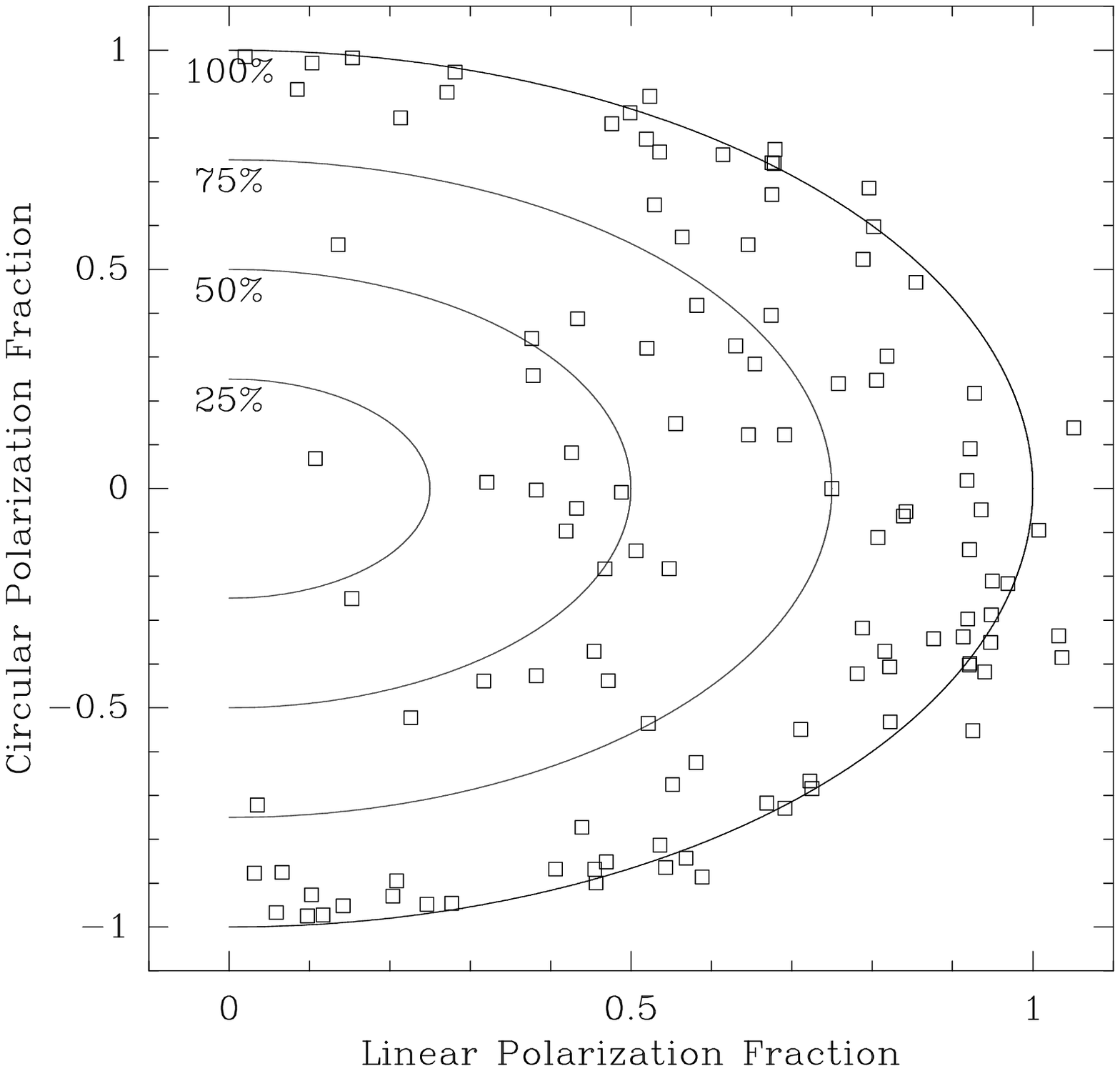}
\end{center}
\singlespace
\caption[Total polarization of maser spots.]{Total polarization of
maser spots.  Only spots with nonzero observed fluxes in all three of
RCP, LCP, and Linear ($\sqrt{Q^2 + U^2}$) are shown, which excludes 942
spots.  Positive circular polarization fraction corresponds to
positive Stokes V (i.e., RCP flux greater than LCP flux).  Curves
showing 25\%, 50\%, 75\%, and 100\% total polarization are drawn.
Most maser spots are $> 75\%$ polarized.
\label{tot-pol}}
\doublespace
\end{figure}

Of the maser spots in our sample, 97\% are at least 75\% polarized.
Although Tables 2 to 20 of Paper I do not list the total polarization
fraction explicitly, it can be approximated by noting that the total
polarization fraction is $\sqrt{Q^2 + U^2 + V^2} / I$, where $Q, U,
V,$ and $I$ are the Stokes parameters.  Stokes $I$ and $V$ can be
obtained from the sum and difference, respectively, of the listed RCP
and LCP flux densities of spots, and the linear flux density gives
$\sqrt{Q^2 + U^2}$.  In 86\% of the maser spots, LCP and RCP fluxes
are not both detected, implying that the spot is circularly
polarized to the limits of detectability.  For other spots,
frequently $Q^2 + U^2 + V^2 \approx I^2$, indicating that many spots
are nearly 100\% polarized, as shown in Figure \ref{tot-pol}.  A
portion of the discrepancy from equality in the above equation can be
explained by a variety of factors.  As can be seen in Tables 2 to 20
of Paper I, when the same maser spot is seen in both RCP and LCP
emission, the position and velocity of the peak emission may be
slightly different in both.  Blending of strong adjacent maser spots
can also make determination of fit parameters difficult.  Furthermore,
the linear polarization fraction of a maser spot may vary across a
spot size.

Nevertheless, there are certainly spots that appear to be only
partially polarized (Figure \ref{tot-pol}).  Unpolarized emission ($Q
= U = V = 0$) would appear as equal flux densities in the LCP and RCP
feeds with no detectable linear polarization.  In W51 e1, for example,
there are a number of maser spots where the LCP and RCP flux densities
are nearly equal and centered at roughly the same position and
velocity.  Since there is nearly no detected linear polarization
in the OH masers in this source, it is possible that these
maser spots are only partially polarized.
These maser spots could also be produced by masing in
sites where the magnetic field splitting is small compared to the line
width \citep{gkk2}.  Our data do not provide sufficient resolution to
tell whether the spots with roughly equal RCP and LCP fluxes are due
to two 100\% polarized spots that are at slightly different positions
and velocities or whether they are due to a single spot that is not
100\% polarized.

\subsection{Overlap of Masing Clumps\label{overlap}}

Theoretically, $\pi$-components are favored for magnetic fields
inclined $\ge 55\degr$ to the line of sight \citep{gkk3,grayfield} and
$\sigma$-components are favored for smaller angles.
Consider an ensemble of maser sites, each threaded by an independent,
randomly-oriented magnetic field.  The fraction of maser sites for
which amplification of $\sigma$-components is favored is
\begin{equation}
2 \cdot \frac{1}{4\pi} \cdot \int_{\phi = 0}^{2\pi}\int_{\theta
= 0}^{55\degr} \sin \theta \, d\theta d\phi = 0.426,
\end{equation}
where the factor of two accounts for the possibilities of
the magnetic field being oriented toward or away from the observer.
Accounting for the fact that two $\sigma$-components are produced for
each $\pi$-component by Zeeman splitting, $\sigma$-components would
comprise $2 * 0.426 / (1 + 0.426) = 60\%$ of total maser spots.

As discussed in \S \ref{magstruc}, the magnetic field in any
individual source is highly ordered, and the distribution of magnetic
field orientations at maser sites is not oriented in a uniformly
random direction.  But for a large sample of sources distributed
throughout the Galactic plane, it is plausible that the range of
magnetic field directions sampled will approach a uniform random
sample.  However, in our sample a large majority of maser spots have a
small or zero linear polarization fraction.  Two-thirds of maser spots
have no detectable linear polarization at all.  As discussed in \S
\ref{faraday}, amplification of $\pi$-components may be reduced by
Faraday rotation along the amplification path.
Two additional effects pointed out by \citet{elitzuriii} may explain
the larger fraction of $\sigma$-components we detect.  First, although
an inclination of $55\degr$ divides angular phase space into two
distinct regions in which $\sigma$- and $\pi$-components dominate, the
relative amplification of $\sigma$- and $\pi$-components is larger for
$\theta < 55\degr$, where $\sigma$-components dominate, than for
$\theta > 55\degr$, where $\pi$-components dominate.  In the
unsaturated regime, the ratio of the absorption coefficients for
$\sigma$- and $\pi$-components, $\kappa^\sigma / \kappa^\pi$, reaches
a minimum of 0.5 at $\theta = 90\degr$.  For comparison,
$\kappa^\sigma / \kappa^\pi$ reaches a value of 2 at $39\degr$ and
grows in an unbounded manner as $\theta \rightarrow 0\degr$ (see \S 3
of \citeauthor{elitzuriii}).  Unsaturated $\pi$-components should on
average be weaker than unsaturated $\sigma$-components, so the
percentage of $\sigma$-components above a reasonable detection
threshold would be higher than the 60\% expected based on an analysis
of the sizes of angular phase space alone.  Second, as masers
saturate, competitive gain will favor the stronger component.
Saturated $\sigma$-components will reduce the absorption coefficient
$\kappa^\pi$ by a factor of 3, but saturated $\pi$-components only
reduce $\kappa^\sigma$ by a factor of 2 \citep{elitzuriii}.  Thus,
$\sigma$-components should be even more numerous than
$\pi$-components even if the masers are saturated.

However, it is clear that some $\pi$-components are seen in the
$^2\Pi_{3/2}, J = 3/2$ maser transitions.  Some maser spots exhibit a
high degree of linear polarization as would be expected for
$\pi$-components, and the distribution of PPA in some sources strongly
suggests that $\pi$-components are seen (see \S \ref{notes-w75n}).
Nevertheless, maser spots that we believe are $\pi$-components are not
100\% linearly polarized.  Since $\Delta m_f = 0$ radiation is
inherently linearly polarized, circular polarization must be generated
externally.  We speculate that these $\pi$-components with nonzero
circular polarization arise from the superposition of two masing
clumps along the line of sight.  If the emission from a
$\pi$-component spot intersects a region of OH appropriately shifted
in velocity, it may stimulate emission in a $\sigma$ mode.  Since the
incident radiation from the first cloud (i.e., the $\pi$-component) is
highly amplified and therefore bright, it can strongly stimulate the
second cloud, since the incident linear polarization will be seen by
the second cloud as a superposition of the two opposite-handed
circular modes.  Even if the amplification in the second cloud is very
weak, a significant amount of circular polarization can be added, and
the radiation will no longer be completely linearly polarized, as
shown in Appendix \ref{stokesappendix}.  The distinction between a
$\pi$- and a $\sigma$-component may be blurred if there is $\gtrsim 1$
gain length of material in the second cloud.

This circularization of a bright $\pi$-component due to an extremely
weak $\sigma$-component requires that the weak cloudlet be in front of
the bright maser from the observer's perspective.  If the
$\sigma$-component is behind the $\pi$-component, the propagation path
of radiation passes first through the weak ($\sigma$) cloudlet and
then through the strong $\pi$-component.  The radiation field that the
cloudlet amplifies, whether background continuum or its own
spontaneous emission, is much weaker than in the case where the
radiation from the $\pi$-component stimulates emission from the
$\sigma$-component, so the superposition of spots would be
indistinguishable from an isolated $\pi$-component with no surrounding
material.

In principle the reverse scenario could occur as well: a
$\sigma$-component stimulates emission in the $\pi$-mode from a
smaller cloud of OH gas at the appropriate velocity for amplification.
This would have the effect of adding extra linear polarization to a
$\sigma$-component.  Since $\sigma$-components are in general
elliptically polarized (i.e., have a nonzero linear polarization
fraction), it may not be possible to distinguish observationally
between a $\sigma$-component that has stimulated weak emission in the
$\pi$-mode from a second maser clump and one that has not.  The linear
polarization fraction of a $\sigma$-component is a function of the
inclination of the magnetic field to the line of sight \citep{gkk2},
so this effect could lead to overestimation of the magnetic field
inclination at OH maser sites.

It is probable that the overlap of maser components along the line of
sight would add circular polarization to $\pi$-components more
systemically than it would add linear polarization to
$\sigma$-components.  For most common bulk material motions (e.g.,
infall, outflow, rotation), the radial component of the velocity field
will change monotonically along a ray from a maser spot to the
observer.  If the change in radial velocity along the line of sight
exceeds the Zeeman splitting between a $\sigma$- and $\pi$-component
(1.2~\kms\ at 1665 MHz and 0.7~\kms\ at 1667 MHz for a 4~mG magnetic
field), the radiation from the $\pi$-component may stimulate weak
amplification from OH in a $\sigma$ mode.  But a $\sigma$-component
could only stimulate emission from the $\pi$ mode of a cloud of OH
along the line of sight if the change in radial velocity were in the
same sense as the Zeeman splitting of the $\sigma$-component.

Given that only a small column density of OH along the line of sight
between a maser and the observer is required to add significant
circular polarization to the observed maser, it is likely that a large
fraction of $\pi$-components will be misidentified as
$\sigma$-components.  Unless there is an abrupt outer edge to the
radial distribution of OH in a massive star-forming region, the
radiation from many $\pi$-components will stimulate weak
emission in a $\sigma$-mode of the surrounding OH.

\subsection{Elongated Arrangements of Maser Spots\label{elongated}}

In several sources, OH masers are found in elongated filamentary
arrangements.  For instance, there is a line of maser spots near the
origin in the W75 S map shown in Figure 25 of Paper I, and the masers in
W75 N (Figure 27 of Paper I) appear to be oriented primarily along two
perpendicular axes.  In the filamentary arrangement of maser spots
seen in the northern grouping of W75 N, the sky projection of the
magnetic field as deduced from the PPA of the linear portion of the
polarization implies that the magnetic field may be aligned
predominantly along the line of elongation.

Often there is a velocity gradient along the elongation, such as in
W75 S or the masers in a NE/SW line in ON~2~N.  These lines of masers
with velocity gradients are common in CH$_3$OH.  \citet{norris93}
observed 10 sources for which the maser spots were distributed mostly
in a line with the major axis of the distribution several times
greater than the minor axis.  Plots of the velocity of the maser spots
versus the major axis offset are generally distributed into two
quadrants rather than tightly along a straight line \citep{norris98}.
The authors speculated that the masers are tracing circumstellar disks
and that the deviation from a straight line in the velocity-major axis
plots is due to maser amplification at different radii in the same
circumstellar disk.

More recent observations do not seem to favor the interpretation of
these maser arrangements as circumstellar disks, however.  First,
\citet{debuizer} looked for H$_2$ $\nu = 1 - 0$ S(1) emission in
massive SFRs for which circumstellar disks were suspected on the basis
of collinear\footnote{We will refer to structures of maser spots
aligned along the same line as \emph{collinear} to avoid any possible
confusion with \emph{linear} polarization.} distributions of methanol
masers.  Since molecular hydrogen is a diagnostic of shocked outflows,
it was expected that the H$_2$ emission would be oriented primarily
perpendicular to the putative disks.  Instead, the H$_2$ emission in
almost all of the sources for which it was detected was preferentially
oriented parallel to the line of methanol masers.  De Buizer suggested
that the masers were instead tracing an outflow.  Second, the proper
motions of maser spots in the two linear structures in G9.62$+$0.20
are directed primarily perpendicular to the structures
\citet{minierevn}.  This suggests that in at least some cases methanol
masers may trace shocks rather than circumstellar disks.  Whether this
applies to similar arrangements of OH masers is not yet well
established.  It is believed that the lifetimes of methanol and
hydroxyl masers overlap but are not identical \citep{reidsumm}.  It is
therefore possible that structures delineated by OH masers trace a
different evolutionary phase of forming high-mass stars than do
CH$_3$OH masers.

An alternate possibility to the disk hypothesis is that these
elongated arrangements of maser spots may simply be a result of the
motion of material threaded by magnetic fields.  This could be a
result of 1) collapse in the early stages of star formation or 2)
shock-driven outflows in later stages.  In case 1), as clouds
of OH and other material fall inward, they will draw the magnetic
field inward with them.  This will tend to align field lines with the
material elongations, provided there is enough angular momentum to
avoid spherical collapse.  In case 2), a shock propagating
outward from the boundary of the UC\hii\ region compresses material
ahead of it, leading to elongations along the shock front.

The drift speed for ambipolar diffusion of a magnetic field out of a
maser cloud is
\begin{equation}
v_\mathrm{d} = 0.6
  \left(\frac{B}{5 \times 10^{-3}~\mathrm{G}}\right)^2
  \left(\frac{n}{10^6~\mathrm{cm}^{-3}}\right)^{-2}
  \left(\frac{r}{10^{15}~\mathrm{cm}}\right)^{-1}
  \left(\frac{\alpha}{2 \times 10^{-9}~\mathrm{cm}^3~\mathrm{s}^{-1}}\right)^{-1}
  \left(\frac{x_e}{10^{-5}}\right)~\mathrm{km}~\mathrm{s}^{-1},
\end{equation}
where $r$ is the radius of a maser spot, $\alpha$ is the ion-neutral
collision rate coefficient, and $x_e$ is the ionization fraction
\citep{black79}.  The drift speed is less than a typical shock speed
($ > 5$~\kms), especially if carbon is ionized in any substantial
fraction.
Thus, the magnetic field will be dragged along and compressed by the
shock, resulting in a field oriented parallel to the shock front and
therefore along the material elongation.  Unless the shock is totally
planar, there may be velocity gradients along the elongations.  The
slight curvature to some of these elongations may also be explained by
the expansion of a spherical shock front or a planar shock into an
inhomogeneous medium.  This is in contrast to models explaining
elongated arrangements of maser spots as disks, where little curvature
would be expected if preferentially seen edge-on.  Nevertheless, some
curvature along the elongations may be produced if a disk is inclined.

\subsection{Is There a Connection Between Maser Clusters and
Shocks?\label{shocks}}

As mentioned in \S \ref{clusterscale}, there is a characteristic maser
clumping scale of $\sim 10^{15}$~cm.  Additionally, maser clusters
tend to be concentrated on or near the periphery of \hii\ regions.
This is not always the case in complicated sources such as W75 N, but
more often than not these maser clusters appear near the boundary of
an \hii\ region, especially given that we observe three-dimensional
distributions of masers in projection.  Occasionally even individual
clusters are elongated in a filamentary manner, as in W3(OH)
\citep{reidw3}.

It is possible that these clusters of OH masers form in shocked
neutral gas outside the ionization boundary.  The initial growth phase
of a UC\hii\ region involves an $R$-type ionization
front \citep[see, e.g.,][]{kahn54}.  When
the speed of the ionization front slows to twice the sound speed of
the ionized material, a transition occurs and the ionization front
changes to a weak $D$-type \citep{shubook}.  This is characterized by
the existence of two separate fronts: an ionization (I) front and a
shock (S) front that precedes it.  \citet{kawamura} directly measured
the expansion speed of the UC\hii\ region, i.e., the I front, in
W3(OH) to be $3 - 5$~\kms.  Since the speed of the I front is less
than the sound speed in the ionized material ($\sim 10$~\kms), the
ionization front must be of $D$-type in W3(OH).

Others have theorized that masers near a UC\hii\ region
appear in the shocked neutral material between the I and S fronts
\citep[e.g.,][]{baldwin,elitzur}.  Theoretical calculations suggest
that I-S fronts are inherently prone to instabilities
\citep{vandervoort}.  Less clear is the exact mechanism of instability
growth, although there is no shortage of candidates \citep{dyson}.
\citet{giuliani} found that the slab of material between the I and S
fronts was unstable to oscillatory transverse perturbations.  The
wavelength of fastest perturbation growth was found to increase with
time.  \citet{vishniac} obtained similar results and additionally
suggested that unstable small-scale fragmentation would eventually
allow neutral gas to be swept up behind the fragments and lead to
Rayleigh-Taylor instabilities, although it is unclear whether this
would occur at the time and size scales of \emph{ultracompact} \hii\
regions.  \citet{pottasch} argued based on the evolutionary sequence
of perturbation shapes of the bright rims in diffuse nebulae that the
Rayleigh-Taylor instability alone cannot be the cause of fingering,
although his analysis was based on older, lower-density \hii\ regions.
Vishniac also speculated that magnetic fields might produce elongated,
filamentary structures.  Two-dimensional simulations by
\citet{garciasegura} demonstrate the appearance of this hydrodynamical
instability regardless of the density structure of the neutral gas and
details of the radiative cooling law included.  The wavelength of the
fastest-growing perturbations increases with time, with interfragment
spacings on the order of several times 100 AU near the base of the
fingers and approximately 100 AU near the tips for UC\hii\ ages near
$10^4$ years, as shown graphically in their Figure 6.  Note that this
is consistent with both the clustering scale and the dynamical age of
OH masers, as discussed in \S\S \ref{clusterscale} and
\ref{relation-hii}.

Assuming that some clusters near UC\hii\ regions form in the shocked
neutral medium between the I and S fronts, two projection-related
factors would imply that clusters should be more frequently found at the
periphery of \hii\ regions rather than atop them, an effect not seen
(see \S \ref{relation-hii}).  First, velocity gradients are likely to
be higher along the fingers of shocked neutral material than across
them.  Amplification lengths should therefore be longer on average for
fingers pointing in the plane of the sky, where the velocity gradient
is tangential in projection, than for fingers pointing toward us
(i.e., projected near the center of the \hii\ region).  Still,
potential path lengths are longer along fingers rather than across
them, and it is unclear whether velocity gradients are large enough to
favor amplification across rather than along the fingers.  Second, the
magnetic field threading the neutral material will be dragged along
with it.  Field lines will be folded such that a field line entering
the finger from the neutral material will exit the finger back into
the neutral material after bending through $\approx 180\degr$.  This
would imply an effective reversal of the line-of-sight direction for
lines of sight along fingers.  However, \emph{intra}-cluster magnetic
field line reversals are never seen.  Indeed, the magnetic fields
deduced from OH maser in clusters superposed atop \hii\ regions in
ON~1 and W3(OH) suggest that there is a consistent line-of-sight field
direction in those sources even atop the \hii\ region.  It is possible
that OH masing clumps occur near the ``palms'' of the fingers or that
the fingers of neutral material containing the OH masers are not very
long.  In either case, the bend in the magnetic field lines could be
much less than $180\degr$, consistent with our lack of detection of
line-of-sight field reversals in these clusters.

Large-scale ($> 1000$~AU) collinear maser structures, such as the
NE/SW line in ON~2~N, probably cannot be explained by shock
instabilities around the UC\hii\ region because they are larger in
scale than the \hii\ region itself.  It is still possible that these
structures occur in neutral gas that has been shocked by another
source in the star-forming complex.  There is often a velocity
gradient along these structures, which could be explained by a curved
or decelerating shock.

\subsection{Relation of OH Masers to Galactic Magnetic Fields}

Noting that line-of-sight directions obtained from Zeeman splitting in
eight Galactic OH maser sources were consistent with a clockwise
Galactic field, \citet{davies} postulated that OH masers in massive
star-forming regions traced the Galactic magnetic field.  Follow-up
studies by \citet{reid90} and \citet{reid93} supported this claim but
were suggestive of a more complicated Galactic field structure.
Subsequent analyses employing this technique on ever-larger sample
sizes have indicated that correlations with the Galactic magnetic
field may exist \citep{baudry97,fish03}, but detailed probing of the
Galactic field with this method remains elusive.  Observational
limitations of this method consist of unknown distances to many of the
sources as well as possibly incorrect magnetic field data resulting
from inadequate spatial resolution to unambiguously identify Zeeman
pairs of maser features, since few sources have been observed at VLBI
resolution.  Additionally, magnetic field information has thus far
been limited to the sign of the line-of-sight field orientation (i.e,
whether the magnetic field points in the hemisphere toward or away
from the Sun).  This may be insufficient to accurately probe a
predominantly toroidal Galactic magnetic field at lower Galactic
longitudes, in the direction of the majority of massive star-forming
regions (as well as most of the spiral structure of the Galaxy).

If the magnetic field orientation in massive star-forming regions is
correlated with the Galactic field, the processes of high-mass star
formation must not tightly wrap the magnetic field configuration
despite the rotation and collapse necessary to produce the central
condensation, a proposition for which there is theoretical support
\citep[e.g.,][]{lishu,allenshu}.  As discussed in \S \ref{magstruc},
an ordered magnetic field can be inferred from the regularity of the
line-of-sight direction of the magnetic field over large portions of
the source and the field strengths inferred from Zeeman pairs within
the same clusters of $10^{15}$~cm.  Furthermore, the magnetic fields
deduced from OH Zeeman splitting in massive SFRs separated by
distances on the order of a kiloparsec show a preference to be
co-aligned \citep{fram}.  A numerical investigation of the collapse of
rotating, magnetized, isothermal cloud cores suggests that collapse
can occur without introducing a significant twist to the magnetic
field \citep{allenshu}.  These authors find that the maximum pitch
angle of the magnetic field is approximately $20 \degr$ along a ridge
of accreting material (see their Figure 4).  Inward from this ridge
the magnetic field resists wrapping, while outward from the ridge the
wrap from differential rotation is small.  Simulations by
\citet{matsumoto04} confirm that the magnetic field of a collapsing
core maintains alignment with the magnetic field of the parent cloud.
They find that a young star's magnetic field is inclined no more than
30\degr\ from that of the parent cloud for a weak initial field
strength ($\approx 20~\mu$G at a density of $2.6 \times
10^4$~cm$^{-3}$), with much better alignment when the initial field
strength is greater.

If these models are correct, the magnetic field orientation before
collapse might be partially preserved in the material surrounding the
core.  Since the fields around newly-formed massive stars are ordered
(see \S \ref{magstruc}), this suggests that magnetic field
orientations deduced from OH maser Zeeman splitting may be indicative
of the Galactic magnetic field.  A VLBI survey of OH
masers in massive star-forming regions would eliminate Zeeman pairing
ambiguity and possibly allow for three-dimensional modelling of the
ambient magnetic field in the few sources in which the Faraday
rotation is small enough that the full magnetic field orientation can
be inferred from the observed linear polarization fraction and PPA of
maser spots.  If accurate distances can be obtained as well, as
through trigonometric parallaxes of higher frequency maser
transitions, OH masers may prove to be a useful tool for probing the
Galactic magnetic field.

\section{Summary of Interstellar OH Maser Properties\label{conclusions}}

\begin{itemize}

\item Ground-state OH masers typically cluster on a scale of $10^{15}$~cm,
providing evidence that their distribution is linked to a process with
an inherent scale, as opposed to turbulence (which is generally
scale-free).  The magnetic field strengths implied by Zeeman splitting
suggest that OH masers occur in regions of density $10^5$ to several
$\times 10^7$~cm$^{-3}$.  OH masers are found preferentially near the
UC\hii\ region in massive SFRs.  Their distribution around UC\hii\
regions suggest an expansion age of $\approx 10^4$ years for typical
expansion velocities.  OH masers do not appear to be systemically
shifted from the velocity of the associated star by more than a few
\kms, although possible exceptions exist, as in G5.886$-$0.393 and
W75~N VLA 2.  Taken together, these pieces of evidence support the
theory that most OH masers occur in the shocked neutral gas between
the ionization and shock fronts of UC\hii\ regions.  The distribution
of maser fluxes with distance from the central UC\hii\ region suggests
that OH masers turn off abruptly rather than weakening gradually after
$\sim 10^4$ years.

\item Some OH masers are seen far from or without any associated \hii\
region.  It is unclear whether these masers are pumped by a star with
an associated weak, undetected hypercompact \hii\ region or whether
they are shock-excited without an ionization front.  In some sources
(e.g., \object[W75S]{W75~S}), OH masers appear to trace a collinear
structure with a velocity gradient.  These formations probably trace
shock fronts rather than protostellar disks.

\item Magnetic fields are ordered in massive SFRs, lending observational
support to theories that indicate that the ambient magnetic field
direction may be preserved during massive star formation.  Nearly all
sources show either a consistent line-of-sight magnetic field
direction or a single reversal of the line-of-sight direction across
the source.  Within a maser cluster of size $10^{15}$~cm,
line-of-sight magnetic field direction reversals are never seen, and
the field strengths deduced from Zeeman splitting are almost always
consistent within $\pm 1$~mG.

\item We do see both $\pi$- and $\sigma$-components, including a ``Zeeman
triplet'' in W75 N (see \S \ref{triplet}).  But OH maser spots that
are 100\% linearly polarized, as theoretically expected of
$\pi$-components, are extremely rare.  There is a range of sources
with qualitatively different linear polarization properties.  At one
extreme (as in W75~N) high linear polarization fractions are seen, and
the PPAs show some correlation with observed structures and probable
magnetic field directions.  In most sources the linear polarization
fractions are much less than 1 and PPAs cannot be easily interpreted
as magnetic field directions.  At the other extreme are sources such
as W51 e1 and e2, in which little or no linear polarization is
detected and the \emph{total} polarization fraction of some maser
spots is much less than unity.

\item The wide range of polarization properties observed in OH masers may be
explained by a combination of Faraday rotation and overlap of maser
components.  If OH masers are indeed near or embedded in \ion{C}{2}
regions, the electron density may be high enough that the masers are
near a critical point of Faraday rotation.  A typical maser spot
likely has large ($> 1$~rad) Faraday rotation over the entire
amplification length, but not over a single gain length of the maser.
If Faraday rotation is a factor of $\sim 5$ lower, the total Faraday
rotation along the amplification path may be small enough such that
the PPAs are still roughly aligned with the magnetic field lines.  On
the other hand, if Faraday rotation is a factor of $\sim 5$ larger,
the Faraday rotation per gain length could exceed 1~rad, destroying
linear polarization and depolarizing the maser.  Even if the Faraday
rotation is small enough to allow amplification of a 100\% linearly
polarized $\pi$-component, its polarization may be partially
circularized by one of the $\sigma$-modes of a weakly-inverted clump
of OH between the maser site and the observer.  This is likely a very
important effect, as only a modest inversion and a small column
density of OH are required to add significant circular polarization to
a $\pi$-component.

\item Theoretically the linear polarization fractions and directions of
maser components can be used to determine the full, three-dimensional
orientation of the magnetic field at masing sites.  But the
interpretation of PPAs may be very difficult in sources for which the
amount of Faraday rotation along the propagation path between the
source and the observer is unknown.  Inferring a magnetic field
orientation in the plane of the sky also requires unambiguous
identification of $\sigma$- and $\pi$-components, due to the $90\degr$
difference in PPA response to a magnetic field.  Zeeman pairs provide
the surest method of identifying $\sigma$-components, but their
polarization properties may be too contaminated by internal Faraday
rotation and maser overlap to permit interpretation of the inclination
of the magnetic field to the line of sight.

\end{itemize}

\acknowledgements

We thank M.~D.~Gray, M.~Elitzur, and J.-P.\ Macquart for helpful
comments in preparation.

{\it Facility: \facility{VLBA}}

\appendix

\section{Notes on Individual Sources\label{sourcenotes}}

\subsection{G5.886$-$0.393\label{appendixgfive}}

There is a reversal of the line-of-sight direction of the magnetic
field across the source.  All Zeeman pairs in the south of the source
indicate a negative magnetic field (i.e., oriented in the hemisphere
pointing toward the Sun), while all pairs in the north of the source
indicate a positive magnetic field.

Nearly all of the maser spots identified in the northeastern cluster
constitute a component of a Zeeman pair.  In total, there are eight
Zeeman pairs in the cluster -- four each in the 1665 and 1667~MHz
transitions.  The magnetic field strengths are consistent, ranging
from 1.2 to 2.0~mG in the cluster.  The center (material) velocities
of the Zeeman pairs range from 8.6 to 10.0~\kms.  This is in excellent
agreement with \citet{caswell01}, who find a 6035 MHz Zeeman pair in
this region centered at 9.96~\kms\ with a splitting of 1.49~mG.

We can define the Zeeman pairing efficiency as twice the number of
Zeeman pairs divided by the total number of maser spots (in both
polarizations and transitions) in a region.  In the limiting case
where every maser spot is a $\sigma$-component in a detectable Zeeman
pair, the Zeeman pairing efficiency would be 100\%.  For the
northeastern cluster (Figure 3 of Paper I), the Zeeman pairing
efficiency is 84\%.  The pairing efficiency in the western half of the
source is only 16\%.  We note that the velocities of maser spots range
from 2.6 to 15.8~\kms\ in the western half of the source This is a
much larger range than for the northeastern cluster, which has a
larger Zeeman pairing efficiency.  Since velocity coherence is
necessary for the amplification of both $\sigma$-components of a
Zeeman pair, it is reassuring to note that the region with the more
coherent velocity field also produces Zeeman pairs more efficiently.

\citet{zijlstra} observe OH emission from $-45$~\kms\ to 17~\kms\ and
interpret this emission as tracing a bipolar outflow.  Our
observations span only the upper end of this velocity range.  We see
redshifted emission extending to the southwest of the western group of
masers, in general agreement with \citeauthor{zijlstra} They do not
see emission at 8.6 to 10.0~\kms\ in the eastern half of the source,
probably due to the large channel width (2.2~\kms\ velocity
equivalent) and beam size of their observations.  They also do not see
emission at the appropriate velocities corresponding to the isolated
$-2.4$ and $+1.2$~mG Zeeman pairs we detect.  At 6035 MHz,
\citet{caswell01} finds emission in the velocity range corresponding
to our observations and absorption from $-25$ to $+2$~\kms, supporting
the model of \citeauthor{zijlstra}

\citet{feldt03} detect a candidate O-type ionizing star at
$\Delta\alpha = -2\farcs 6 \pm 0\farcs 2, \Delta\delta = 0\farcs 0 \pm
0\farcs 2$ in Figure 1 of Paper I.  The star's location is coincident with
the isolated $1.2$~mG Zeeman pair to within positional errors.  The
brighter RCP component of this Zeeman pair has a flux density of
2.37~Jy.  Since the \hii\ region is optically thick at $\lambda =
18$~cm \citep{afflerbach96}, it is unlikely that direct amplification
of stellar radiation is important.
Is the spatial coincidence of this maser spot with the projected
location of the star due to chance, or is amplification favored due to
the stellar radiation?
Nine other main-line maser
features in G5.886$-$0.393 are brighter than this feature, indicating
that the observed flux density is not strongly affected by whether the
maser is projected atop the star.  However, the observed flux density
of strong ($> 1$~Jy) maser spots may be relatively insensitive to
initial conditions if they are at least partially saturated.  Due to
interstellar scattering, it is possible to get only a lower limit for
maser brightness temperatures, but it is probable that the brightest
maser spots in this source are saturated (see \S \ref{saturation}).

\subsection{G9.622$+$0.195}

The G9.62$+$0.19 complex contains several UC\hii\ regions, as well as
a hot molecular core \citep{cesaroni}.  We detect OH masers around
sources D, E, and G in the nomenclature system of \citet{garayg9} and
\citet{testi}.  At source E, we find two Zeeman pairs which both
indicate a positive magnetic field.  At source G, we find two Zeeman
pairs which both indicate a negative magnetic field.  Few OH maser
spots and no Zeeman pairs are identified at source D, while no maser
activity at all is seen near the hot core (source F), located between
sources D and G.

We do not see the isolated 1665 MHz RCP maser detected between sources
E and G in \citet{arm}.  This maser spot has likely weakened below our
detectability threshold in the 10 years since their observations.
Water maser emission is seen at this site and associated with sources
D, E, and G \citep{hofner96}.

\subsection{G10.624$-$0.385}

Overall evidence indicates that
\object[G10.624-0.385]{G10.624$-$0.385} is undergoing collapse.
Observations of NH$_3$ show that the molecular material is rotating
and falling inward, with rapid spiral motions inward of about 0.05 pc
\citep{ho86,keto87}.  The plane of rotation cuts through the center of
the \hii\ region and is oriented approximately 20\degr\ west of north
\citep{keto88}.  Coherent inward motions are also seen within the
ionized gas in the \hii\ region \citep{keto02}.

We detect relatively few maser spots in G10.624$-$0.385.  Emission
falls into three regions: a clump to the east, a clump $2\arcsec$ to
the west of the previous clump, and an isolated maser spot to the
northwest.  We find only one Zeeman pair, indicating a magnetic field
of $-6$~mG in the easternmost clump.  The west clump appears arclike,
with a length of about 150~mas.  This is the only clump to show any
linear polarization, most of which is oriented roughly perpendicular
to the arc.  None of the maser spots we detect lies along the plane of
rotation.

The ground-state masers we detect span a velocity range of $-2.4$ to
$+3.3$~\kms.  This is in agreement with the velocity span of excited
states of OH: $-2.0$ to $-1.5$~\kms\ in $^2\Pi_{1/2}, J = 1/2$
emission \citep{gardner83}, $-2.5$ to $+1.0$~\kms\ in $^2\Pi_{3/2}, J
= 7/2$ absorption \citep{gbtoh}, and $-1.9$ to $-0.4$~\kms\ in
$^2\Pi_{3/2}, J = 9/2$ absorption \citep{walmsley86}.

\subsection{G34.257$+$0.154}

The G34.3$+$0.2 complex contains several UC\hii\ regions.  Most
prominent is the ``cometary'' \hii\ region, labelled C in the
nomenclature of \citet{gaume94}.  The morphology of region C can be
explained as a bow shock due to the supersonic relative motion of the
exciting source with respect to the surrounding medium
\citep{reid85,maclow91}.  Region C may be composed of more than one
continuum component \citep{sewilo04}.  Two fainter components,
labelled A and B by \citet{reid85}, are located to the southeast and
northeast of region C, respectively.  The complex has a kinematic
distance of 3.8~kpc (Galactic center distance $r_0 = 10$~kpc)
\citep{reifenstein}.

A blueshifted outflow extending to the northwest is seen by
\citet{hatchell01}.  Components A and C were detected in the
mid-infrared, but component B was not, suggesting that it is deeply
embedded and therefore very young \citep{campbell00}.  Based on this
and the spectral index of component B \citep[$0.9 \pm 0.4$,
from][]{gaume94}, Hatchell et al.~conclude that component B is the
source of the outflow.

Unlike other sources, there is not a single line that can be drawn
across the entirety of G34.257$+$0.154 which separates regions of
positive and negative magnetic field.  However, such lines can be
drawn for the masers associated with regions B and C separately.
Region B contains many Zeeman pairs with positive magnetic field near
the continuum source and a single Zeeman pair with negative magnetic
field to the northeast of the source.  All of the Zeeman pairs ahead
of the bow shock in region C imply a negative magnetic field except
for a small region near the \hii\ region to the south.  There is good
qualitative agreement between the magnetic field measured here and
similar VLBA observations of the source in 1995 by \citet{zrm}.  In
the northern half of the source, the field directions are identical
between the two observations, and field magnitudes agree to better
than 0.3~mG where they overlap.  In the center and south, we find
larger field magnitudes than reported by Zheng et al.  Line-of-sight
field directions are again in agreement, although Zheng et al.\ find
Zeeman pairs implying magnetic fields of $+0.5$ and $-0.5$~mG near the
origin of Figure 9 of Paper I.  At this same location, \citet{gasiprong}
find only one Zeeman pair implying a magnetic field of $-5.0$~mG.  In
this region, we infer magnetic fields of $-0.6$, $-5.1$, $-5.7$, and
$-6.0$~mG from four Zeeman pairs.  At 6035 MHz, \citet{caswellv} and
\citet{caswell01} find three Zeeman pairs indicating a magnetic field
of $-4$~mG.  Their observations do not have the necessary angular
resolution to determine which \hii\ region the 6035 MHz are associated
with, but their results are largely consistent with magnetic fields
obtained from the cometary region C, especially in the north.

\citet{gasiprong} also observed linear polarization with the MERLIN
array.  It is difficult to compare our results directly with theirs,
since the $0\farcs 16$ resolution afforded by MERLIN is insufficient
to separate distinct maser components with very different linear
polarization position angles.  Nevertheless, for the brighter spots it
is possible to identify a maser spot in our data that corresponds to a
similar spot in the Gasiprong study.  Polarization position angles
generally agree to 10$\degr$ or $20\degr$.

\subsection{G35.577$-$0.029}

This source contains two UC\hii\ regions.  All of the maser activity
appears to be associated with the western \hii\ region.  We detect
maser spots only on the western limb of the western \hii\ region with
the exception of one isolated spot to the east.  Due to registration
uncertainties, it is not clear whether this spot is located directly
atop the \hii\ region or on the eastern limb.  We detect three Zeeman
pairs in the western clump of emission, with all three implying a
magnetic field in the range of $-4$ to $-6.3$~mG.

\subsection{G40.622$-$0.137}

There is a large ($\sim 1\arcsec$) cluster of OH maser emission
centered approximately $1\farcs 5$ away from the only detected \hii\
region.  We detect two Zeeman pairs in the same region, consistent
with a field strength of approximately $-6$~mG.  \citet{caswellv} find
one Zeeman pair in 6035~MHz OH maser emission implying a magnetic
field of $+1.7$~mG, suggesting that there may be a reversal of the
line-of-sight direction of the magnetic field in this source.
Methanol and water masers are also seen within less than 1\arcsec\ of
the reference position \citep{forster89,beuther02}.

\subsection{G43.796$-$0.127\label{notes-g43}}

The X-band continuum maps show two sources -- a bright source to the
northwest, and a weak source to the southeast.  All maser emission
lies atop the northwest source.  A total of seven Zeeman pairs were
identified.  Five of these indicate a positive magnetic field, and two
indicate a negative magnetic field.  Unlike in other sources in which
a reversal is seen, it is not possible to draw a single straight line
such that the magnetic field on each side of the line has a uniform
line-of-sight direction.  A Zeeman measurement of 6035~MHz
(excited-state) OH emission implies a magnetic field of $+3.6$~mG
\citep{caswellv}.  Three measurements of the Zeeman effect in H$_2$O
masers, which trace a higher range of densities, imply a magnetic
field of $-13.3$ to $-46.1$~mG \citep{sarma}.

We have adopted an ammonia velocity of 45.2 \kms\ for G43.796$-$0.127.
The NH$_3$ spectrum itself (see Figure 41 of Paper I) is complicated, and
it is difficult to identify which line is the main line and which are
hyperfine lines.  We identify the line at 45.2 \kms\ as the main line
because its velocity most closely matches that of the CS $J = 7
\rightarrow 6$ velocity of 44.3 \kms\ \citep{plume}.  Since the critical
density of CS $J = 7 \rightarrow 6$ is $2 \times 10^7$ cm$^{-3}$, we
feel confident that it is tracing the same high-density material as
the NH$_3$ emission.

\subsection{W51}

We find a total of 46 Zeeman pairs near sources e1 and e2, making W51
the most prolific massive SFR in terms of the number of Zeeman pairs
in our survey.  As previously reported \citep{arm2}, source e2
contains two Zeeman pairs implying the strongest magnetic fields ever
seen in interstellar OH masers: 19.8 mG and 21 mG.

Source e1 shows a reversal of the line-of-sight field direction, which
points toward the Sun in the northern half of the source and away from
the Sun in the southern half.  Source e2 is the clearest example yet
that shows the extent to which magnetic fields are ordered in massive
SFRs.  All 22 Zeeman pairs indicate a positive magnetic field.
Although there is a huge variation in the strength of the magnetic
field across the source, multiple Zeeman pairs in each cluster have
consistent field strengths to within about 1 mG.

W51 is remarkable among our source sample as having almost no
detectable linear polarization.  Three maser spots near the origin in
source e1 have linear polarization fractions of 1\% to 2\%.  No linear
polarization was detected for any other spot in source e1 or any spot
at all in source e2.  The plausibility of circularization due to high
Faraday rotation along maser amplification paths is discussed in \S
\ref{faraday}.

\subsection{ON~1\label{notes-on1}}

The maser spots appeared to be grouped into three regions.  The masers
in the northern group have velocities near 4~\kms.  The masers in the
central group are located at about $13-14$~\kms.  The southern maser
spots fall primarily along an extended collinear feature.  The
velocities in this line range from 13 to 15~\kms.  OH masers are not
seen at intermediate velocities.  Methanol masers show a similar
velocity structure \citep{szymczak00}.

Zeeman splitting in ON~1 is everywhere consistent with a magnetic
field pointing in the hemisphere toward the Sun.  We do not find any
unambiguous Zeeman pairs in the northern group.  At 6031 and 6035 MHz,
\citet{desmurs98} find four Zeeman pairs implying magnetic fields from
$-3.6$ to $-6.3$~mG.  These maser spots appear to fall slightly north
and west of the northern group of maser spots detected in the
ground-state transitions here and in \citet{arm}, but their center
velocities fall between 13.7 and 15.3~\kms\ as compared with a
velocity range of 3.2 to 6.2~\kms\ in the northern ground-state group.
It is unclear whether this 10~\kms\ difference reflects a large
velocity gradient in the northern part of the source or whether the
6031 and 6035 MHz emission comes from a different area, reflecting
registration uncertainties between the various sets of observations.
Two 13441 MHz Zeeman pairs at 14.1 and 0.3~\kms\ indicate magnetic
fields of $-3.8$ and $-8.3$~mG, respectively \citep{gbtoh}.

Since registration uncertainties of a few tenths of an arcsecond may
exist between the continuum image and our maser spot maps, one
possible interpretation is that the northern and southern maser groups
are located on the limb of the UC\hii\ region, while the center group
is projected onto the \hii\ region.  While the precise locations of
these groups relative to the \hii\ region along the line of sight is
unknown, the center group must be located in front of the \hii\
region, since the \hii\ region is optically thick at 18~cm
\citep{zhengon1}.  The authors also noted an arcminute-scale gradient
of $11 \pm 2$~\kms~pc$^{-1}$ in NH$_3$ emission with a velocity of
$\sim 12$~\kms\ near the \hii\ region (comparable to that of the
southern group of OH masers) and an H$76\alpha$ recombination velocity
of $5.1 \pm 2.5$~\kms, blueshifted with respect to the NH$_3$
emission.  They concluded that the motions in ON~1 were consistent
with infall and rotation.

The velocities of the three groups of maser spots we observe may be
consistent with infall and rotation on a smaller angular scale as
well.  If the north and south maser groups are at limbs of the
rotation,
the implied rotation speed would be 5~\kms\ centered at 9~\kms\ and
roughly aligned with the direction of rotation noted by Zheng et al.
The center maser group might then be infalling at 5~\kms\ as well.  At
this radius, 5~\kms\ corresponds to the freefall velocity for a
$20~M_\sun$ star, so if net rotation is also sustained, rapidly
spiralling infall must be occurring.  Nevertheless, this could explain
why the RCP maser emission in the northern group is seen farther from
the center of the \hii\ region than the LCP emission.  Because the
magnetic field splits the RCP emission to a lower LSR velocity than
the LCP emission, the coherent path length is larger farther away from
the center of the \hii\ region.

\subsection{K3$-$50}

This \hii\ region has a diameter of over 0.1~pc, which is large for an
ultracompact \hii\ region.  Masers are found only to the north and
east of the \hii\ region.  The line-of-sight magnetic field direction
points toward the Sun at all maser groups.  However, due to the lack
of maser emission to the south and west of the \hii\ region, we cannot
conclusively rule out a magnetic field reversal across \object[NAME
K3-50A]{K3$-$50}.  Using the Effelsberg 100 m telescope,
\citet{baudry97} find two 6035 MHz Zeeman pairs indicating field
strengths of $-5.3$ and $-9.1$~mG centered at $-18.68$ and
$-19.44$~\kms, respectively.  These are redshifted compared to the
Zeeman pairs we identify, whose center velocities range from $-22.30$
to $-19.79$~\kms.  Two components at $-20.1$ to $-20.2$~\kms\ and
$-25.0$ to $-25.5$~\kms\ are seen in absorption in the $^2\Pi_{3/2}, J
= 7/2$ lines \citep{gbtoh}.

\subsection{ON~2~N}

Eleven Zeeman pairs have been identified in ON~2~N, all indicating a
positive magnetic field.  All of the OH maser emission is located to
the south and west of the UC\hii\ region, in the same area as the
H$_2$O maser emission \citep{hofner96}.

There are three groups of maser spots arranged roughly in a line with
a position angle 35\degr\ east of north beginning at the \hii\ region.
This line of spots exhibits a velocity gradient with the most
redshifted emission toward the southwest.  A cluster of maser spots
with a large velocity dispersion is offset from this line and
elongated perpendicular to it.  Almost all spots with any detectable
linear polarization, as well as all spots with a high linear
polarization fraction, are offset from the line of masers.

\subsection{W75 S}

We have identified 13 Zeeman pairs around the UC\hii\ region in
W75 S.  There is a reversal of the line-of-sight magnetic field
direction across this SFR.  Seven Zeeman pairs to the east indicate a
magnetic field pointing toward the Sun, while six to the west
indicate a magnetic field in the opposite direction.  Additionally,
field strengths within maser clusters are remarkably consistent.  The
six Zeeman pairs in the western cluster imply field strengths of
$5.6$ to $6.6$~mG, and the six Zeeman pairs to the southeast of the
\hii\ region imply field strengths of $-3.8$ to $-5.3$~mG.

\begin{figure}
\begin{center}
\includegraphics[height=3.0truein]{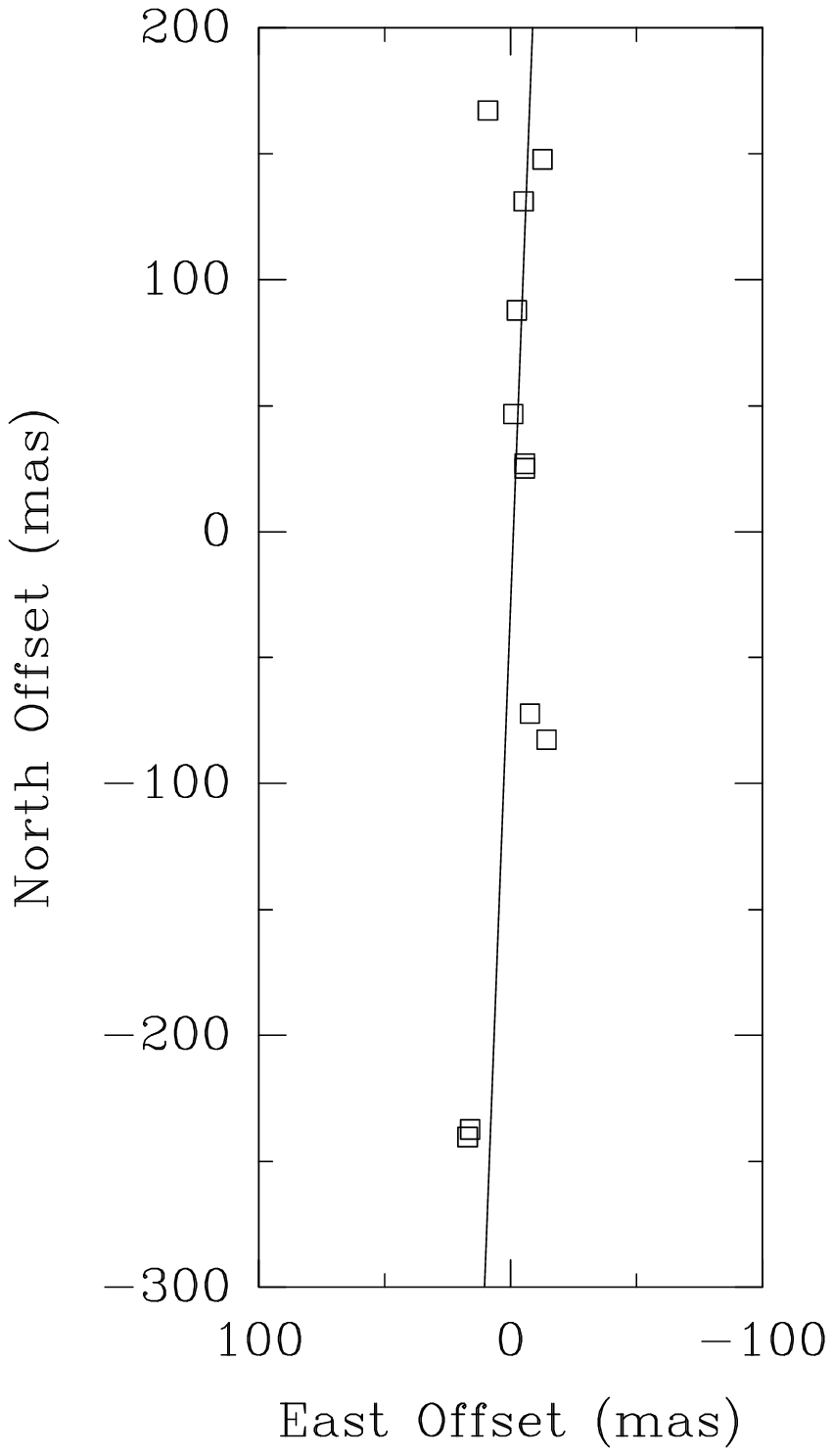}
\includegraphics[height=2.5truein]{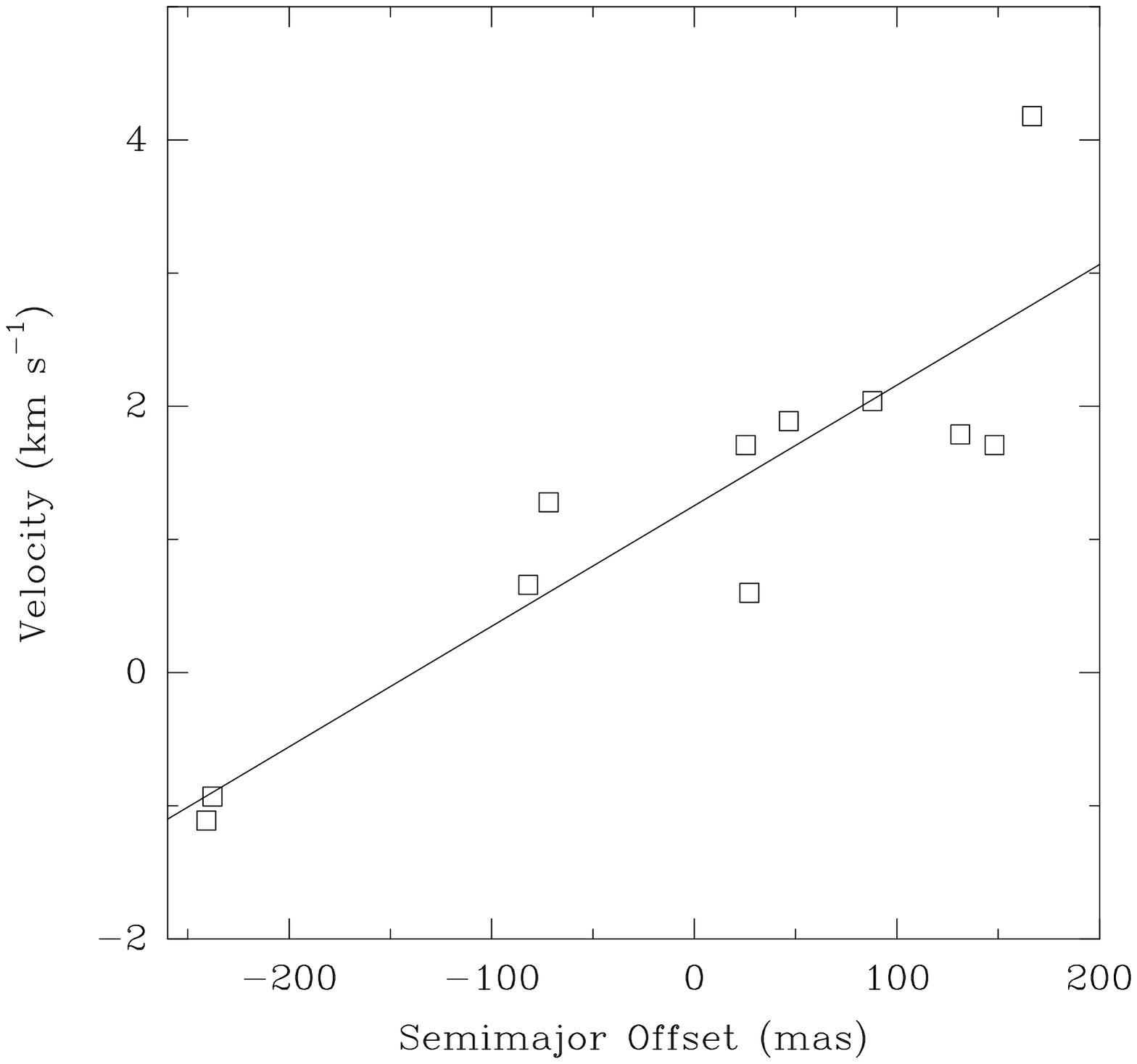}
\end{center}
\singlespace
\caption[Collinear maser arrangement in W75 S]{Left: Enlargement of
collinear maser arrangement in W75 S.
The maser spots are tightly grouped along a line, as
shown.  Right: Plot of radial velocity $v_\mathrm{LSR}$ versus
distance along major axis.  The line of best fit, representing a
velocity gradient, is shown.
\notetoeditor{These are two separate files, so they can be
  rescaled and positioned independently.}
\label{w75sdisk}}
\doublespace
\end{figure}

A collinear arrangement of maser spots exists near the origin in
Figure 25 of Paper I.  An enlargement of this region is shown in Figure
\ref{w75sdisk}, along with a best-fit line.  There is a velocity
gradient along this line with the velocity increasing to the north.
If these masers are interpreted as tracing a circumstellar disk in
Keplerian rotation, the radius of the disk is at least 400~AU, and the
central mass is at least 6~M$_\sun$.  See \S \ref{elongated} for
further discussion of the possible interpretations of this structure.

\subsection{W75 N\label{notes-w75n}}

The region of OH masing in W75 N coincides with three continuum
sources, identified from north to south as VLA~1, VLA~2, and VLA~3 by
\citet{torrelles}.  Based on an elongation of VLA~1 (at position angle
$\sim 43\degr$) and the spectral index of the source, the authors
conclude that there is an ionized, biconical, partially optically
thick jet emanating from the source.  In addition to this flow, there
is a larger, 3-pc flow oriented at position angle $62 \fdg 5$ that
does not appear to be driven by the outflow in VLA~1 \citep{shepherd2}.

W75 N contains OH maser spots distributed primarily along two axes.
Along the north-south axis the magnetic field is oriented away from
the Sun, while along the east-west axis the field is oriented toward
the Sun.  At the intersection of these two axes there are two Zeeman
pairs, each indicating a different sign of the magnetic field.  It
appears that most masers are associated with VLA~1.  However, the
dynamics of the cluster at the limb of VLA~2 (Figure 28 of Paper I)
are unlike the rest of the masers, suggesting that these masers are
indeed associated with VLA~2.

Other maser species exist in W75 N as well.  H$_2$O masers are seen
close to the \hii\ regions, with spots distinctly on the limbs of
VLA~2 and VLA~3 as well as along the position angle of the jet in
VLA~1 \citep{torrelles}.  CH$_3$OH masers are seen primarily as
an extension to the north of the north-south axis of OH masers, and
they are distributed in a line with position angle $42\degr$
\citep{minier2}.  Thus, the CH$_3$OH masers appear to be associated
with VLA~1.

The region around VLA~2 contains a large number of OH maser spots at a
wide range of velocities.  The difference in velocity between the most
blueshifted and most redshifted spot in this region is 34 \kms\
\citep{ellder,hutawarakorn}.  Our bandwidth only covered about 21
\kms\ of this range.  We detect maser emission in both the highest and
lowest usable velocity channel at which we observed.  A large velocity
dispersion in H$_2$O masers is also seen in this region
\citep{torrelles2}.  Based on these large velocity dispersions, the
location of VLA~2, and the steeply rising spectrum of continuum
emission, \citet{hutawarakorn} conclude that VLA~2 is the source of
the large-scale molecular outflow.  \citet{slysh01} interpret the
maser spots in the north-south axis as being a disk centered at VLA~1.

Since the observations detailed in Paper I, a 1665 MHz maser in W75 N
has become the strongest ever detected, reaching a flux of
approximately 1 kJy \citep{alakoz05}.  They find a 750 Jy RCP feature
at 1.8 \kms, as well as two other new features near 0 and $-1$ \kms.
All three features are predominantly linearly polarized and are
therefore likely $\pi$-components or $\sigma$-components where the
magnetic field is oriented close to the plane of the sky.  These are
offset by about $-0.5$ \kms\ compared to the bright features in Paper
I.  It is possible that these features are new or that the masers near
VLA~2 are accelerating.  This latter possibility cannot be ruled out
because of the nature of the masers near VLA~2.  They are observed to
span a velocity range 34 \kms\ wide and appear to be associated with
an outflow \citep{ellder,hutawarakorn}.  If the outflow is
decelerating, it is possible that masers entrained in the flow may
appear at slightly different velocities between epochs.

\subsection{Cep A}

Cep A is a complex molecular cloud condensation.  \citet{hughes84}
detected no fewer than 14 \hii\ regions in the complex, and subsequent
observations have uncovered even more radio continuum sources
\citep{hughes88,curiel02}.  HW 2, the brightest continuum source in
Figure 30 of Paper I, appears to contain at least four compact sources
\citep{hughes95} and is believed to be the source of two thermal jets
\citep{rodriguez94,hughes01}.  The 6 cm radio jet is oriented at a
position angle of 44$\degr$ and is observed to have a projected
velocity of $950 \pm 150$ \kms\ \citep{rodriguez01}.  Water masers are
seen associated with this jet \citep{torrelles96}, and their proper
motions suggest the presence of at least three distinct sites of star
formation within a projected 200 AU radius \citep{torrelles01}.  To
the south in Figure 30 of Paper I are three continuum sources: HW 3c,
HW 3div, and HW 3dii in the nomenclature of \citet{hughes95} and
\citet{torrelles98}.  Source HW 3c shows evidence of multiple
components \citep{hughes95}.  Water masers are seen around HW 3dii and
the nearby source HW 3di, which is not detectable in our X-band image
\citep{torrelles98}.  We detect OH masers around HW 2, HW 3c, and HW
3div, as well as a cluster of masers between HW 3div and HW 3dii and a
lone maser not near any continuum source.

\citet{bartkiewicz05} identify seven Zeeman pairs at 1665 MHz and two
at 1667 MHz with MERLIN.  Four of these (Z$_2$, both Z$_7$'s, and
Z$_8$) agree with our findings in terms of central velocity and
magnetic field strength to within the errors expected from velocity
resolution.  Pairs Z$_5$ and Z$_6$ also agree with the magnetic field
strength we find in the respective maser clusters, although several
maser spots are blended together at these locations in the MERLIN
beam.  We do not find counterparts for pairs Z$_1$ and Z$_3$, and we
do not have the velocity coverage necessary to observe Z$_4$.

\subsection{NGC 7538}

\object{NGC 7538} is a complex star-forming region.  The continuum
source in Figure 32 of Paper I, known as IRS 1, contains a core of two
compact components and a larger, spherical region to the south
\citep{turner84, campbell84}.  \citet{scoville86} argued for the
existence of an ionized stellar wind outflow based on the spectral
index of millimeter continuum emission and for a possible disk
oriented east-west based on $^{13}$CO emission.  \citet{gaume95} find
that the distribution of continuum emission is clumpy and suggest that
photoionization from the central star is responsible for this
emission.  There is a bipolar, high-velocity CO flow around IRS 1
\citep{fischer85}, possibly collimated by a denser ring of material
seen in CS \citep{kawabe92}.  This outflow, as well as other outflows
and stellar winds in the IRS 1-3 region, may be driving the expansion
of a molecular half-shell \citep{xu03}.

NGC 7538 is also remarkable in the variety of maser species detected.
Among these are rare maser species such as H$_2$CO \citep{forster80},
$^{14}$NH$_3$ \citep{madden86}, and $^{15}$NH$_3$
\citep{mauersberger86}.  H$_2$O masers \citep{genzel76,kameya90} and
CH$_3$OH masers from a variety of transitions
\citep[e.g.,][]{wilson84,wilson85,batrla87} are seen in NGC 7538 as
well.  OH maser emission has been seen in the 1665, 1667, and 1720 MHz
$^2\Pi_{3/2}, J = 3/2$ lines
\citep[e.g.,][]{downes70,dickel82,hutawarakorn03}, the 6035 MHz
$^2\Pi_{3/2}, J = 5/2$ line \citep{guilloteau84}, and the 4765 MHz
$^2\Pi_{1/2}, J = 1/2$ line \citep{palmer84}.
\citeauthor{hutawarakorn03} find one Zeeman pair in each of the 1667
and 1720 MHz transitions implying magnetic fields of $-1.7$ and
$-2.0$~mG, respectively.  We find no Zeeman pairs at 1667 MHz, but we
do detect a Zeeman pair of $+0.7$~mG at 1665 MHz, suggesting that
there is a reversal of the line-of-sight direction of the magnetic
field across the source.

\subsection{S269}

We have found three Zeeman pairs in S269, consistent with a magnetic
field of $-4.0$ to $-4.5$~mG.  Otherwise, S269 is one of the simplest
sources in our study.  It exhibits few maser spots.  There is not much
linear polarization of maser emission in this source.  The VLA survey
was unable to detect any continuum emission, nor was ammonia emission
found.  And the range of velocities of maser emission is a mere
$3.9$~\kms.

\begin{figure}
\begin{center}
\includegraphics[width=6.0in]{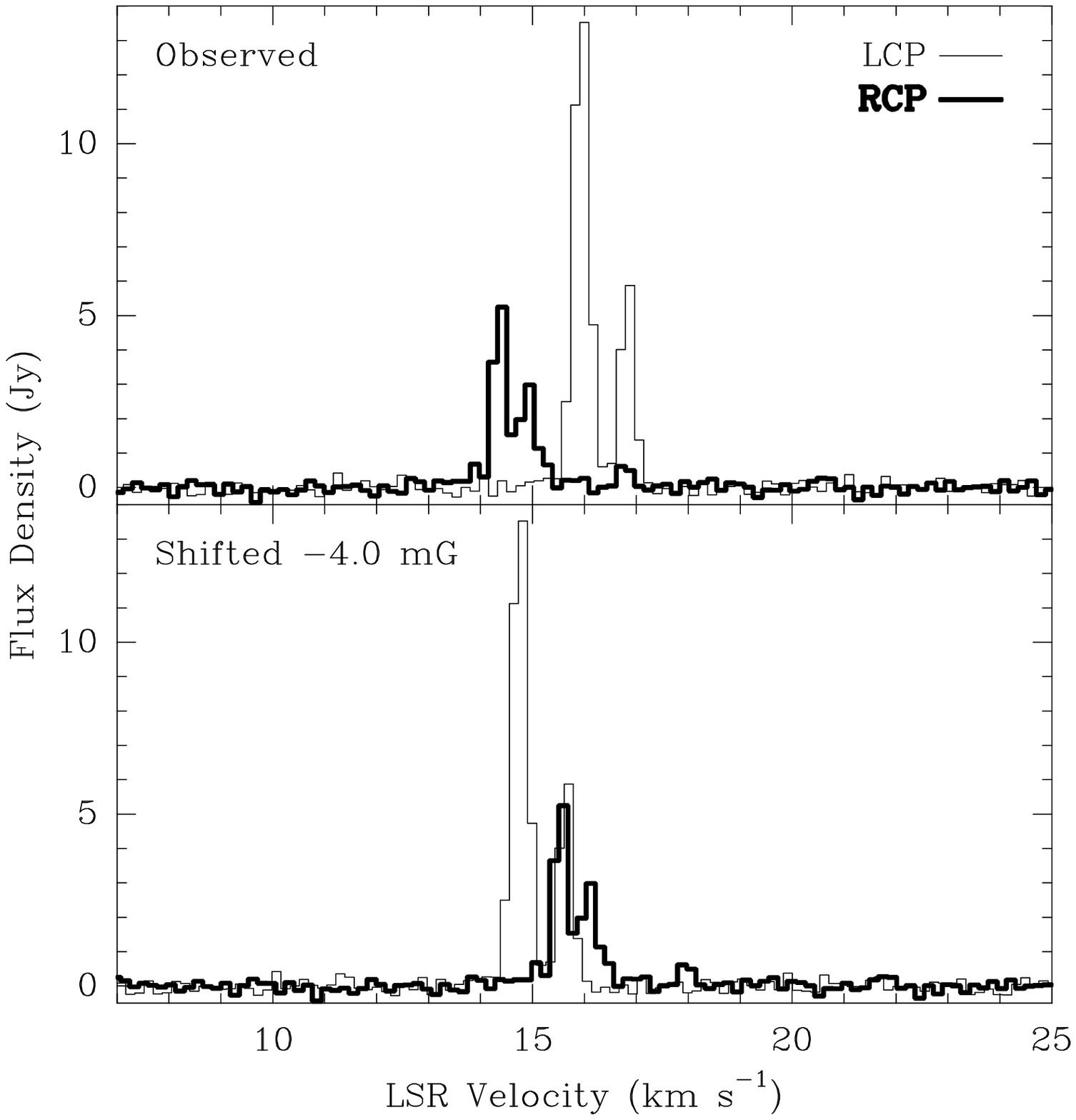}
\end{center}
\singlespace
\caption{Top: Observed spectra of 1665 MHz emission from S269.  RCP
  emission is shown in bold, and LCP emission is shown in normal
  weight.  The two weak RCP bumps at the same velocities as the strong
  LCP features at 16.0 and 16.8 \kms\ are due to the elliptical
  polarization of the LCP features.  Bottom: The same spectra when
  corrected for a $-4.0$ mG magnetic field.  Note that the velocity
  range is more than halved compared to the observed
  spectra.\label{s269-spec}}
\doublespace
\end{figure}

The magnetic field appears to be oriented toward the Sun everywhere
across the region of maser emission.  All RCP emission, excluding weak
features associated with the linearly-polarized component of two
strong left-elliptically polarized masers, occurs at lower velocity
than the LCP emission.  Figure \ref{s269-spec} shows the spectrum of
the 1665 MHz emission.  When the spectra are corrected for Zeeman
splitting (``demagnetized'') for a $-4.0$ mG magnetic field, the total
velocity range spanned by the RCP and LCP features decreases from 3.9
\kms\ to 1.5 \kms.

S269 also exhibits a high degree of variability.  \citet{clegg93}
notes a 1665 MHz maser of 16 Jy LCP flux at 17.9 \kms\ in 1991.5,
diminishing to 1 Jy by 1992.1.  In 1991.6, \citet{arm} find that this
maser has a flux of 7.5 Jy LCP.  The closest feature we find to
matching this is a 1665 MHz maser of 0.11 Jy LCP flux at 17.8 \kms.
We detect this feature in the 17.76 and 17.93 \kms\ velocity channels,
but the lack of detection in a third channel prevents us from being
able to accurately determine the velocity and linewidth of this
feature.

\subsection{Mon R2}

The Mon R2 molecular cloud is one of the nearest high-mass
star-forming regions, but it contains almost no stars of spectral type
earlier than B1 \citep{hughes85}.  An unusual property of Mon R2 is
that maser emission from the 4765 MHz ($^2\Pi_{1/2}, J = 1/2, F = 1
\rarr 0$) transition of OH is stronger than from the ground-state
($^2\Pi_{3/2}, J = 3/2$) set of transitions, suggesting that the
physical conditions are denser and hotter than normally seen for
ground-state OH masers, as noted by \citet{smits98}.  They detect two
masers at 10.65~\kms\ with linear polarization position angles of 13
and 14\degr.  This is in excellent agreement with the brightest maser
feature we detect, coincident with the brightest 4765 MHz maser to
within registration uncertainties, which has linear polarization with
a position angle of 13\degr.  The velocity of this 1665 MHz maser
feature is 10.29~\kms\ when corrected for the Zeeman splitting of the
$-2.6$~mG magnetic field at the site.  The 4765 MHz masers in Mon R2
are highly variable, doubling in strength in less than 19 days and
reaching a peak of nearly 80 Jy before ``disappearing''
\citep{smits03}.  \citeauthor{smits03} finds that 1665 and 1667 MHz
emission is much less variable, varying relatively smoothly with
changes in flux density not exceeding a factor of two over a timescale
of more than four years.

\subsection{G351.775$-$0.538}

G351.775$-$0.538 contains what was the strongest known interstellar OH
maser spot (400 Jy in LCP) until the recent flare in W75 N
\citep{alakoz05}.  \citet{ch80} first noted that the brightness of
this maser spot is highly variable, and it has been monitored
frequently since then \citep[see ][]{macleod}.  Ground-state masers
have previously been seen at velocities as low as $-27.8$~\kms\
\citep{arm} and as high as $7$~\kms\ \citep{macleod}.  Our VLBA
observations covered only the top half of this range.

There is a reversal of the line-of-sight magnetic field direction
across the source, as has been previously noted at 1665 and 1667 MHz
by \citet{arm2} and \citet{fram}.  This reversal is seen at 1720 MHz
as well, where \citet{caswell04} finds magnetic fields of $+3$ and
$-6$~mG.  \citet{caswellv} find a $-3.3$~mG field at 6035 MHz.

Because G351.775$-$0.538 is a low Declination source, $(u,v)$-coverage
is poor, especially along north-south baselines.  The synthesized beam
and spot sizes as listed in Table 20 of Paper I are thus very
large.  This may explain the separations of Zeeman components for
G351.775$-$0.538 (see Table 21 of Paper I), which are larger than
for other sources.

\section{Maser Overlap Polarization Calculation\label{stokesappendix}}

In this appendix we consider the polarization properties expected of a
$\pi$-component that stimulates weak amplification from a
$\sigma$-mode of a clump of OH between the first maser and the
observer, as described in \S \ref{overlap}.  It is helpful to analyze
the radiation in terms of the Stokes parameters, which are defined in
terms of the electric fields in the radiation as follows:
\begin{eqnarray}
I &=& <\epsilon_x \epsilon_x^*> + <\epsilon_y  \epsilon_y^*> \nonumber \\
Q &=& <\epsilon_x \epsilon_x^*> - <\epsilon_y  \epsilon_y^*> \nonumber \\
U &=& <\epsilon_x \epsilon_y^*> + <\epsilon_x^*  \epsilon_y> \nonumber \\
V &=& i(<\epsilon_x \epsilon_y^*> - <\epsilon_x^*  \epsilon_y>)
\end{eqnarray}
They can also be written in terms of the electric fields in the two
senses of circular polarization:
\begin{eqnarray}
I &=& \frac{1}{2}(I_{rr} + I_{ll}) \nonumber \\
Q &=& \frac{1}{2}(I_{rl} + I_{lr}) \nonumber \\
U &=& \frac{i}{2}(I_{rl} - I_{lr}) \nonumber \\
V &=& \frac{1}{2}(I_{rr} - I_{ll}), \label{stokesdef2}
\end{eqnarray}
where $I_{rr} = <\epsilon_r \epsilon_r^{*}>$, etc.

Consider a simple case of weak maser amplification in the absence
of Faraday rotation.  We will start with radiation that is 100\%
linearly polarized in the $x$-direction, as could be produced by a
$\pi$-component.  The Stokes parameters of the radiation are
\begin{equation}
I = I_0, Q = I_0, U = 0, V = 0,
\end{equation}
so the polarization fractions ($m_L$ linear, $m_C$ circular, and $m_T$
total) are
\begin{equation}
m_L = 1, m_C = 0, m_T = 1.
\end{equation}
Now suppose that this radiation is fed into a second, weak maser spot
shifted in velocity such that emission is stimulated in the
RCP $\sigma$-mode.  The amplification factor is such that the flux
density in the RCP mode is multiplied by a factor of 2.  Then
\begin{eqnarray}
I_{rr} &\rightarrow& 2I_{rr} = 2I_0 \nonumber \\
I_{ll} &\rightarrow& I_{ll} = I_0   \nonumber \\
I_{rl} &\rightarrow& \sqrt{2} I_0   \nonumber \\
I_{lr} &\rightarrow& \sqrt{2} I_0,
\end{eqnarray}
where the factors of $\sqrt{2}$ are due to the fact that if the flux
density increases by a factor of 2, the electric field amplitude
increases by a factor of $\sqrt{2}$.  Substituting these values into
equation (\ref{stokesdef2}) results in the following:
\begin{eqnarray}
I &=& \frac{1}{2}(2I_0 + I_0) = \frac{3}{2} I_0 \nonumber \\
Q &=& \frac{1}{2}(\sqrt{2} I_0 + \sqrt{2} I_0) = \sqrt{2} I_0
  \nonumber \\
U &=& \frac{i}{2}(\sqrt{2} I_0 - \sqrt{2} I_0) = 0 \nonumber \\
V &=& \frac{1}{2}(2I_0 - I_0) = \frac{1}{2} I_0. \label{stokesresults}
\end{eqnarray}
Converting equation (\ref{stokesresults}) to polarization fractions,
we obtain
\begin{eqnarray}
m_L &=& \frac{\sqrt{Q^2+U^2}}{I} = \frac{2\sqrt{2}}{3} \approx 0.943
  \nonumber \\
m_C &=& \frac{V}{I} = \frac{1}{3} \approx 0.333
  \nonumber \\
m_T &=& \frac{\sqrt{Q^2+U^2+V^2}}{I} = \sqrt{m_L^2+m_C^2} = 1.000.
\end{eqnarray}
The net effect is that the maser is still 100\% polarized, but the
linear polarization fraction has dropped to less than unity and the
circular polarization fraction is reasonably large.  For amplification
of a single circular mode by a factor of $n$, the linear polarization
fraction is
\begin{equation}
m_L = \frac{2\sqrt{n}}{1+n}.
\end{equation}
The circular and linear polarization fractions become equal at $n
\approx 6$, or approximately 1.8 $e$-fold amplification lengths for an
unsaturated maser.

{}

\end{document}